\documentclass[12pt]{article}
\usepackage{wider}
\usepackage{color}
\usepackage{comment}
\usepackage{cite}
\usepackage{fnpct}

\begin{document}

\newcommand{\sqvb}{\ensuremath{ \langle \!\langle 0 |} }
\newcommand{\sqvk}{\ensuremath{ | 0 \rangle \!\rangle } }
\newcommand{\sqvn}{\ensuremath{ \langle \! \langle 0 |  0 \rangle \! \rangle} }
\newcommand{\wh}{\ensuremath{\widehat}}
\newcommand{\be}{\begin{equation}}
\newcommand{\ee}{\end{equation}}
\newcommand{\bea}{\begin{eqnarray}}
\newcommand{\eea}{\end{eqnarray}}
\newcommand{\ra}{\ensuremath{\rangle}}
\newcommand{\la}{\ensuremath{\langle}}
\newcommand{\rra}{\ensuremath{ \rangle \! \rangle }}
\newcommand{\lla}{\ensuremath{ \langle \! \langle }}
\newcommand{\str}{\rule[-.125cm]{0cm}{.5cm}}
\newcommand{\pr}{\ensuremath{^{\;\prime}}}
\newcommand{\ppr}{\ensuremath{^{\;\prime \prime}}}
\newcommand{\da}{\ensuremath{^\dag}}
\newcommand{\as}{^\ast}
\newcommand{\eps}{\ensuremath{\epsilon}}
\newcommand{\ve}{\ensuremath{\vec}}
\newcommand{\ka}{\kappa}
\newcommand{\non}{\ensuremath{\nonumber}}
\newcommand{\lf}{\ensuremath{\left}}
\newcommand{\rt}{\ensuremath{\right}}
\newcommand{\al}{\ensuremath{\alpha}}
\newcommand{\dfn}{\ensuremath{\equiv}}
\newcommand{\ga}{\ensuremath{\gamma}}
\newcommand{\ti}{\ensuremath{\tilde}}
\newcommand{\wti}{\ensuremath{\widetilde}}
\newcommand{\hs}{\ensuremath{\hspace*{.5cm}}}
\newcommand{\bet}{\ensuremath{\beta}}
\newcommand{\om}{\ensuremath{\omega}}
\newcommand{\kp}{\ensuremath{\kappa}}

\newcommand{\cO}{\ensuremath{{\cal O}}}
\newcommand{\cS}{\ensuremath{{\cal S}}}
\newcommand{\cF}{\ensuremath{{\cal F}}}
\newcommand{\cX}{\ensuremath{{\cal X}}}
\newcommand{\cZ}{\ensuremath{{\cal Z}}}
\newcommand{\cG}{\ensuremath{{\cal G}}}
\newcommand{\cR}{\ensuremath{{\cal R}}}
\newcommand{\cV}{\ensuremath{{\cal V}}}
\newcommand{\cC}{\ensuremath{{\cal C}}}
\newcommand{\cP}{\ensuremath{{\cal P}}}
\newcommand{\cH}{\ensuremath{{\cal H}}}
\newcommand{\cN}{\ensuremath{{\cal N}}}
\newcommand{\cE}{\ensuremath{{\cal E}}}

\newcommand{\pup}{\ensuremath{^{(p)}}}
\newcommand{\prpr}{\ensuremath{\prime \prime }}

\newcommand{\hsp}{\ensuremath{\hspace*{5mm} }}
\newcommand{\sbp}{\ensuremath{_{[p]} }}

\newcommand{\erf}{\mathop{\mathrm{erf}}}

%To see labels in DVI:
%\newcommand{\xxx}[1]{\hspace*{\fill}{\bf #1}\\}
%\newcommand{\yyy}[1]{\hspace*{\fill}{\bf #1}\\}
%\newcommand{\zzz}[1]{{\bf #1}\\}
%\newcommand{\xyz}[1]{{\bf #1}}
%NB No underscore allowed in name.
%To hide labels in DVI:
\newcommand{\xxx}[1]{}
\newcommand{\yyy}[1]{}
\newcommand{\zzz}[1]{}
\newcommand{\xyz}[1]{}

\title{\bf Probability, Preclusion and Biological 
Evolution in Heisenberg-Picture Everett  Quantum Mechanics
}
\author{
Mark A. Rubin\\
\mbox{}\\
markallenrubin@yahoo.com\\
\mbox{}\\
}
%\date{\mbox{}}
\maketitle
%\clearpage
\begin{abstract}
The fact that certain 
``extraordinary'' probabilistic phenomena---in particular, macroscopic violations of the second  law of thermodynamics---have never been  observed to occur can be accounted for  by taking hard preclusion 
as a basic physical law; i.e. precluding from existence events
corresponding to very small but nonzero values of quantum-mechanical weight.  This approach is not consistent
with the usual ontology of the Everett interpretation, in which outcomes correspond to branches of the  state vector, but can
be successfully implemented using a Heisenberg-picture-based ontology in which outcomes are encoded 
in transformations of operators. Hard preclusion can provide an explanation for biological evolution, which can in turn explain our subjective experiences of, and reactions to, ``ordinary'' probabilistic phenomena, and the compatibility of those experiences and reactions with what we conventionally take to be objective probabilities arising from physical laws.

\mbox{}


\end{abstract}

\section{A categorization of probabilistic phenomena}\label{SecIntro}
\xxx{SecIntro}

The issue of  how probability can either be incorporated  into, or be shown to arise from, Everett quantum mechanics \cite{Everett57,DeWittGraham73,Price95,Barrett99,Vaidman02,HewittHorsman09,Saundersetal10,Wallace12,Wallace13} has long been contentious    (\cite{Everett57,DeWitt72,Graham73, Okhuwa93, Kent90,Deutsch99,Barnumetal00,Rae09,Saunders10,Papineau10,Wallace10,GreavesMyrvold10,Kent10,Albert10,Price10,Zurek10,Schack10,Vaidman11,SebensCarroll18} are but a sampling of the extensive and diverse literature on the subject).
It has been argued that at least some of the 
difficulty associated with this issue is related to difficulty in understanding the nature of probability {\em per se}\/ \cite{Papineau10,Wallace13}.
So it makes sense to focus directly on the  physical phenomena that we associate with probability. These can usefully be placed into two categories:
\begin{description}
\item[Extraordinary probabilistic phenomena]  These are events to which we ascribe  extremely low probability.  Not ``low probability'' at the level of winning a lottery; rather, ``low probabiiilty'' at the level of macroscopic entropy decrease, e.g, observing an ice cube placed in water becoming colder while the water becomes warmer.  The  characteristics of extraordinary probabilistic phenomena  relevant for the present discussion are twofold:
\begin{description}
\item[\rm \em Nonobservation] Although physical theory ascribes to them a nonzero probability of occurring, they have in fact never been observed to occur.  
\item [\rm \em Collective nature] The smallness of the probability of occurrence is related to the fact that the phenomena in question involve a large number (order of Avogadros's number) of constituents (typically molecules).
\end{description}
\item[Ordinary probabilistic phenomena] All other phenomena of physics which are described using probability: passage of photons through a polarizing beam-splitter, absorption of a photon during photosynthesis, radioactive decay of a uranium nucleus\ldots
\end{description}

To start, we focus on extraordinary probabilistic phenomena and, making use of the nonprobabilistic parts of the quantum formalism,  attempt to account for their nonobservation.\footnote{We are of course not claiming that there is any difficulty within the usual theory of probability in understanding  why events with extremely low but nonzero probability have not been observed; after all, we have observed until now only a finite number of events.  The issue is that, as detailed in the above-cited references, there are difficulties in incorporating the usual theory of probability into Everett quantum mechanics. The approach we are taking is to start with the formalism of Everett quantum mechanics, which is deterministic, modify it in a way that preserves this determinism, and show that the resulting theory successfully accounts for the various kinds of phenomena which in, non-Everettian versions of quantum mechanics, are described probabilistically.}
{\em Preclusion}\/, an approach applied to the Everett interpretation in a ``soft'' qualitative form by Geroch   \cite{Geroch84} and presented in a ``hard'' quantitative form by Buniy, Hsu and Zee (BHZ)\cite{BHZ06}\footnote{Having been unaware of this important work at the time of writing of my previous paper on preclusion in Everett quantum mechanics\cite{Rubin10}, I  regrettably failed to cite it there.}, suggests itself.\footnote{Preclusion plays a role in the non-Everett approaches due to Sorkin \cite{Sorkin94,Sorkin97,Sorkin07a,Sorkin07b,DowkerGhaziTabatabai08b,DowkerGhaziTabatabai08a,Barnettetal07,Craigetal07,Ghazi-TabatabaiWallden09a,Ghazi-Tabatabai09,Ghazi-TabatabaiWallden09b,Gudder09,Gudder11,FraucaSorkin17} and  Galvan  \cite{Galvan07a,Galvan07b,Galvan08a,Galvan08b}.  Hanson \cite{Hanson03,Hanson06}  employs the related idea of ``mangling'' in the context of the world-counting approach to probability in Everett quantum mechanics.  Cournot's principle \cite{Cournot1843,ShaferVovk06}, essentially soft preclusion, is used in the theory of classical probability as axiomatized by Kolmogorov \cite[p.4]{Kolmogorov56} to relate the mathematical formalism to physical reality.
 ``An event with very small probability is {\em morally impossible}\/:  it will not happen\cite{ShaferVovk06}.'' }  
BHZ propose that components of the Schr\"{o}dinger-picture wavefunction with norm less than a small positive number be
removed from the wavefunction.   In other words, events with quantum-mechanical amplitudes smaller than a certain level simply do not occur.

This would seem to explain, by fiat, the nonobservation of extraordinary probabilistic phenomena.  The probability of an event in the standard Copenhagen interpretation of quantum mechanics is given by the Born rule; that is, it is the square norm of the component of the wavefunction corresponding to that event. Removing from the wavefunction  events with sufficiently small norm  therefore implies the nonoccurrence of those events. 

However, the fact that the events we are most interested in precluding are collective events presents a difficulty if we employ the usual ontology of the Everett interpretation in the Schr\"{o}dinger picture.  

In Sec.~\ref{SecIncompatibility} below we describe this difficulty in detail.  Sec.~\ref{SecEQMHP} reviews the Heisenberg-picture description of ideal measurement in Everett quantum mechanics and the ontology that it motivates.  In Sec.~\ref{SecModel} we show how the preclusion of extraordinary probabilistic phenomena can be successfully implemented in the Heisenberg-picture Everett interpretation.  
Sec.~\ref{SecPreclusionConsCon} details a consistency condition that preclusion should obey to provide sensible descriptions of sequences of measurements in which the outcomes of later measurements are contingent on the outcomes of earlier measurements.  In 
Sec.~\ref{SecOrdProbEventsBiolEvol} we show how  preclusion of extraordinary probabilistic phenomena provides an explanation for biological evolution and, in particular, for the evolution of subjective judgements regarding ordinary probabilistic phenomena, present a preliminary model of biological evolution relying on preclusion rather than objective probability, and review recent experiments pointing to  the  evolutionary origins of subjective probability.  We conclude with a summary of our results (Sec.~\ref{SecSumm}) and a discussion (Sec.~\ref{SecDisc}).

\section{Incompatibility of preclusion and Schr\"{o}dinger-picture Everett ontology}\label{SecIncompatibility}
\xxx{SecIncompatibility}

We take as a model collective phenomenon  the Bernoulli process and the associated law of large numbers.  The classical Bernoulli process is a sequence of $N$\/ binary events with probability $p$\/ for one of the  two possible outcomes, say outcome $1$\/, (and for outcome $2$\/,  of course, probability $q=1-p$\/).  The law of large numbers, a theorem which follows from the axioms of probability theory, states that by taking  $N$\/ sufficiently large, the probability that the relative frequency $f_1$\/ for outcome $1$\/ (i.e., fraction of events with outcome $1$\/ ) differs by any fixed amount from $p$\/ can be made as small as desired.  Equivalently, the probability that $f_1$\/ is within a fixed amount of $p$\/  can be made as close to unity as desired.   The probability distribution for  $f_1$\/ thus becomes peaked around the value $p$\/.  What is key in deriving this result is that the probability distribution is for relative frequency {\em regardless of order}\/; i.e. it is the sum of the probabilities for all sequences with a given relative frequency.

The quantum analog of the Bernoulli process involves a product state $|\Psi\rangle$\/ of $N$\/ qubits or spins (we will use the two terms interchangeably), i.e., quantum systems each of which has a state space spanned by two basis vectors $|{\cal S}^{(i)};\alpha_1\rangle$\/, $|{\cal S}^{(i)};\alpha_2\rangle$\/, with the amplitudes associated with the basis vectors the same for each qubit. That is,
\begin{equation}
|\Psi\rangle=\prod_{i=1}^N(c_1|{\cal S}^{(i)};\alpha_1\rangle+c_2|{\cal S}^{(i)};\alpha_2\rangle)\label{productstate}
\end{equation}
\xxx{productstate}
where $\alpha_1$\/, $\alpha_2$\/ are nondegenerate eigenvalues labeling the basis vectors and
\begin{equation}
|c_1|^2+|c_2|^2=1.\label{normalization}
\end{equation}
\xxx{normalization}

This can be expanded as
%\be
%%                  |\Psi\rangle=\sum_{b=1}^{2^N}|B_b\rangle
%                  |\Psi\rangle=\sum_{i_1=1}^2\sum_{i_2=1}^2 \ldots \sum_{i_N=1}^2 |B_{i_1,i_2, \ldots i_N}\rangle \label{branches}
%\ee
\be
%                  |\Psi\rangle=\sum_{b=1}^{2^N}|B_b\rangle
                  |\Psi\rangle=\sum_{i_1=1}^2 \ldots \sum_{i_N=1}^2 |B_{i_1, \ldots i_N}\rangle \label{branches}
\ee\xxx{branches}
where

\be
|B_{i_1,\ldots i_N}\rangle=%\sum_{i^{(1)}=1}^2\ldots\sum_{i^{(N)}=1}^2 
c_1^{r_1(i^{(1)},\ldots,i^{(N)})}
c_2^{r_2(i^{(1)},\ldots,i^{(N)})}
%|\cO;\bet_{i^{(1)}},\ldots,\bet_{i^{(N)}}\ra
\prod_{p=1}^N |\cS\pup;\al_{i^{(p)}}\ra,
\label{SPEvmeas}
\ee
\xxx{SPEvmeas}
is one of the $2^N$\/ {\em branches}\/ of $|\Psi\rangle$\/, and where
\be
r_i(i^{(1)},\ldots,i^{(N)})=\sum_{p=1}^N \delta_{i,i\pup},\label{ridef}
\ee
\xxx{ridef}
satisfying
\be
r_1(i^{(1)},\ldots,i^{(N)})+r_2(i^{(1)},\ldots,i^{(N)})=N,\label{rsum}
\ee
\xxx{rsum}
is the number of factors of $c_i$ in each term in (\ref{productstate}). 
Each branch of the state vector    
is regarded as an Everett  world in which the corresponding sequence of measurement outcomes
$\al_{i^{(1)}},\ldots, \al_{i^{(N)}}$\/ has occurred. 
 The number of  $\alpha_1$\/ 
 outcomes in the world  with outcome sequence $\al_{i^{(1)}},\ldots, \al_{i^{(N)}}$\/  
is  $r_1(i^{(1)},\ldots,i^{(N)})$\/  and the relative frequency
of $\alpha_1$\/   
outcomes in this world is
\be
f_1(i^{(1)},\ldots,i^{(N)})=r_1(i^{(1)},\ldots,i^{(N)})/N.\label{f1}
\ee
\xxx{f1}

For sufficiently large $N$\/ we would  like to preclude all branches except those for which the relative frequency  $f_1(i^{(1)},\ldots,i^{(N)})$\/
is close to the Born-rule value 
\be
f_{B,1}=|c_1|^2.\label{Bornrule1}
\ee
\xxx{Bornrule1}
The weight (square norm) of a branch  with outcome sequence $\al_{i^{(1)}},\ldots, \al_{i^{(N)}}$\/     is given by
\be
w(i^{(1)},\ldots,i^{(N)})=
|c_1|^{2r_1(i^{(1)},\ldots,i^{(N)})} |c_2|^{2r_2(i^{(1)},\ldots,i^{(N)})}.\label{Bornrule2}
\ee
\xxx{Bornrule2} 
The weight of a branch with the Born-rule
relative-frequency of ``spin up'' ($\alpha_1$\/) is
\be
w_B=|c_1|^{2N|c_1|^2}\left(1-|c_1|^2\right)^{N\left(1-|c_1|^2\right)}\label{wB}
\ee
\xxx{wB}
So, to establish the Born rule by preclusion,  we  have to find a real number, $\eps_P$\/, smaller than $w_B$\/ and 
larger that the weight of any other branch.\footnote{Of course in general there will be no branch with precisely the relative frequency (\ref{Bornrule1}), because   $N|c_1|^2$\/  may not be an integer.  The goal then would be preclude all branches except the one with the weight closest to $w_B$\/.}

 But this is in general  impossible. Suppose  $|c_1| > |c_2|$\/. 
Then the weight (\ref{Bornrule2}) of
a branch will be a monotonically-increasing function of   $r_1(i^{(1)},\ldots,i^{(N)})$\/,  the number of  $\alpha_1$\/
outcomes in that branch. In particular, the branch with $(i^{(1)},\ldots,i^{(N)})=
(1,1,1,\ldots,1)$\/ will have larger weight than any other, contradicting the requirement that
the Born-rule weight is larger.  We obtain the same type of contradiction in the case $|c_2| > |c_1|$\/. 
For $|c_1|=|c_2|=1/\sqrt{2}$\/ all branches have the same weight. It is only if 
$|c_1|=1$\/ or  $|c_2|=1$\/ that a preclusion approach can work.

If branches with common values of relative frequency are grouped together prior to being
subject to preclusion, then indeed the law of large numbers with the Born-rule value for the peak relative frequency can be recovered using
the same mathematics as in the classical-probability case. Groups of branches with relative frequencies near the
Born-rule value (\ref{Bornrule1}) survive preclusion essentially because there are so many branches in the group, even
though there are branches with non-Born values of relative frequency which individually have larger weights.
For details  see BHZ.

But what is the justification for grouping together branches in this manner before applying preclusion? What dictates that we should group together
branches with the same relative frequency $f$\/ for outcome 1, even though, e.g., one branch may
have the {\em first}\/ $f N$\/ components with outcome 1 and another may have the {\em  last} $f N$\/ components with outcome 1? Might we not just as well consider that, say, a branch with the first $1.01f N$\/ components with outcome 1 should be grouped together with a branch with the first $fN$\/ components with outcome 1?
Keep in mind that the components correspond to measurements of  distinguishable and possibly macroscopic systems.

This, then, is the issue: The usual ontology of the Everett interpretation
regards 
individual Everett branches as worlds.
But it is the {\em sums}\/  of weights for branches with nearby relative frequencies which obey Bernoulli's
law of large numbers.  
If we want to make use of 
preclusion to give physical meaning to 
the branch weights in a way that will  yield the law of large numbers with a peak at the Born-rule value,  we require  an ontology which focuses on 
entities which effectively correspond to the required groups of
branches.

The same  issue of grouping also arises in attempts to demonstate the existence of probability and the Born rule
in Everett quantum mechanics  in the context of infinite ensembles\cite{DeWitt72,Okhuwa93}. 
In that context I have demonstrated that using an ontology motivated
by the  Heisenberg-picture  formulation of the Everett interpretation  and taking into account the fundamental
physical limitations on relative-frequency-measuring devices yields the Born rule in
a straightforward manner\cite{Rubin03}. This approach also works in conjunction with finite
ensembles and preclusion, as  will now be  shown.\footnote{The methodology used in the calculations which appear in Sec.~\ref{SecModel} below differs somewhat from that employed in the corresponding section, 3.5, of \cite{Rubin03}, and allows us to extend some of the results of the latter; see footnote \ref{footnotenumber}, p. \pageref{footnotepage}. }

\section{Ideal measurement in Everett quantum mechanics in the Heisenberg picture}\label{SecEQMHP}
\xxx{SecEQMHP}

We cast our description of the quantum Bernoulli process in terms of observers  performing ideal measurements on the ensemble of two-state systems.  In this section we will describe ideal measurement of a single system, and in the next section progress to measurements of ensembles of systems and relative frequency.
%In the Schr\"{o}dinger picture measurement of \ldots 
%
%To see that this is the case, 
%

The Schr\"{o}dinger-picture description of ideal measurement    
\cite{dEspagnat76}  of a single system $\cS$\/  involves
%Schr\"{o}dinger-picture schematic of measurement in the Everett interpretation,
an observer $\cO$\/ equipped with a measuring apparatus  who measures $\cS$\/  in a superposition of eigenstates of the
observable which the apparatus measures, and in doing so evolves from a state of ignorance to a state of entanglement with $\cS$\/:
%correlated superposition:
\be
|\cO;\bet_0\ra \left(c_1|\cS;\al_1\ra + c_2|\cS;\al_2\ra\right) \rightarrow
c_1|\cO;\bet_1\ra|\cS;\al_1\ra + c_2|\cO;\bet_2\ra|\cS;\al_2\ra.
\label{qmeas}
\ee
\xxx{qmeas}
In the Heisenberg picture 
the state vector does not change, so  we must look at changes
in the operators to be able to characterize ideal measurement.
% time evolution in a manner
%corresponding to the Schr\"{o}diger-picture relation (\ref{qmeas}).
The time evolution in  (\ref{qmeas}) can be  impemented by  
\be
\wh{U}=\exp(-i\wh{H}(t-t_{in})), \label{U_iHt}\label{UintermsofH}
\ee
\xxx{UintermsofH}
where $t$\/ is the time at the end of the measurement,  $t_{in}$\/ is the
time at the beginning (which we also take to be the initial time at which
the Heisenberg and Schr\"{o}dinger pictures coincide), and where
\be
\wh{H}=\sum_{i=1}^2 \wh{h}_i^{\cO} \otimes \wh{P}_i^{\cS}, \label{Hdef}
\ee
\xxx{Hdef}
with
\be
\wh{h}_i^{\cO}=i\kappa (|\cO;\bet_i\ra\la \cO;\bet_0|-|\cO;\bet_0\ra\la \cO;\bet_i|),
\hspace*{5mm}i=1,2,
\label{hdef}
\ee 
\xxx{hdef}
%leads via (\ref{uform}) to $\wh{u}_i^{\cO}$\/'s satisfying (\ref{orthou}),
%provided
%\be
%\kappa=\frac{\pi}{2(t-t_{in})}.)
%\ee
and
\be
\wh{P}_i^{\cS}= |\cS;\al_i\ra\la \cS;\al_i|. \label{PSdef}
\ee
\xxx{PSdef}

 Let $\wh{a}$\/ and $\wh{b}$\/  be the Schr\"{o}dinger-picture operators of which $|\cS;\al_i\ra$\/, $|\cO;\bet_i\ra$\/  in (\ref{qmeas})
are the respective eigenvectors,
\be
\wh{a}=\sum_{i=1}^2\al_i\wh{P}^{\cS}_i,\label{adef}
\ee
\xxx{adef}
\be
\wh{b}=\sum_{i=0}^2\bet_i\wh{P}^{\cO}_i,\label{bdef}
\ee
\xxx{bdef}
with
\be
\wh{P}_i^{\cO}= |\cO;\bet_i\ra\la \cO;\bet_i|. \label{POdef}
\ee
\xxx{POdef}
We will take the $\bet_i$\/'s to be nondegenerate,
and take $t$\/ to be equal to the time at which  a measurement begun at time $t_{in}$\/
is complete, i.e., 
\be
t=t_{in}+\frac{\pi}{2\kappa}.\label{tval}
\ee
\xxx{tval}
Then the operator corresponding to the spin state of the systems  $\cal{S}$\/\/ is
unchanged by the measurement interaction:
%Using (\ref{HPevolution}), (\ref{initopsequal})   and (\ref{UintermsofH})-(\ref{tval}),
\be
\wh{a}(t)=\wh{a}\label{aconst}
\ee
\xxx{aconst}
(we distinguish Heisenberg-picture operators by explicit time arguments),
while the operator $\wh{b}$\/ corrresponding to the state of awareness
of the observer $\cal{O}$\/ is changed  by the measurement interaction into the
form
\be
\wh{b}(t)=\sum_{i=1}^2\wh{b}_i\otimes\wh{P}_i^{\cS},\label{bt}
\ee
\xxx{bt}
where
\be
\wh{b}_i=\exp(i \wh{h}_i^{\cO}(t-t_{in}))\wh{b}\exp(-i \wh{h}_i^{\cO}(t-t_{in})).
\label{biafter}
\ee
\xxx{biafter}
%\be
%\wh{u}_i^{\cO}= \exp(-i \wh{h}_i^{\cO}(t-t_{in})),  \label{uform}
%\ee
%\xxx{uform}
%so
%\be
%\wh{u}_i^{\cO}|\cO;\bet_0\ra=|\cO;\bet_i\ra, \hspace*{.5cm}i=1,2. \label{orthou}
%\ee
%\xxx{orthou}
For the time-independent  Heisenberg-picture state vector we take the  before-measurement
state vector on the left-hand side of (\ref{qmeas}),
\be
|\psi_{in}\ra=|\cO;\bet_0\ra \left(c_1|\cS;\al_1\ra + c_2|\cS;\al_2\ra\right).\label{psiindef}
\ee
\xxx{psiindef}

We find that in the Heisenberg picture  this ideal measurement has the following characteristics \cite{Rubin01,Rubin03,Rubin04, Rubin06}:
\begin{description}
\item[]{\em Before measurement} (time $t_{in}$\/).
 
The time-independent Heisenberg-picture state vector is an eigenstate of the time-dependent Heisenberg-picture operator corresponding to the observer's state of awareness, 
\be
\wh{b}(t_{in})=\wh{b},
\label{bbefore}
\ee
\xxx{bbefore}
with the eigenvalue
corresponding to a state of ignorance:
\be
%\wh{b}|\cO;\bet_0\ra=\bet_0|\cO;\bet_0\ra,\label{bactoin}
\wh{b}|\psi_{in}\ra=\bet_0|\psi_{in}\ra.\label{baction}
\ee
\xxx{baction}
\item[]{\em After measurement} (time $t$\/).

The operator corresponding to the observer's state of awareness has changed from the form
(\ref{bbefore}) to the form (\ref{bt}), a sum of operators $\wh{b}_i$\/ which act
in the state space of the observer  $\cal{O}$\/ multiplied by projection  operators into states of the observed system $\cal{S}$\/. The time-independent Heisenberg-picture state vector is an eigenstate of the $\wh{b}_i$\/'s with eigenvalues corresponding to states in which the observer has
observed the $\cal{S}$\/ in the state labeled by eigenvalue $\al_i$\/:
%(\ref{tval}) and (\ref{uform})
\be
%\wh{b}_i|\cO;\bet_0\ra=\bet_i|\cO;\bet_0\ra.\hspace*{5mm}i=1,2.\label{biaction}
\wh{b}_i|\psi_{in}\ra=\bet_i|\psi_{in}\ra.\hspace*{5mm}i=1,2\label{biaction}
\ee
\xxx{biaction}

\end{description}

%So far
Up to this point we have only transcribed the description of ideal measurement from the Schr\"{o}dinger
picture to the Heisenberg picture. However, since in the Heisenberg picture it is the operators
that evolve, it is natural to think in terms of an ontology reflecting this difference \cite{Rubin03}. Rather than  states (Everett branches)  which
encode properties of an entire  globally-defined world at  a given time, and  in which
properties such as spin or states of awareness of observers have definite values when
the state is an eigenstate of a suitable operator,  we regard the things which exist in the multiverse \cite{Deutsch02} as correponding
not to states but to operators, which have definite values when they satisfy ``eigenoperator'' conditions
such as  (\ref{baction}), (\ref{biaction}).  During the ideal-measurement process, the state of awareness of
$\cO$\/ undergoes a transition from a single entity in a state of ignorance ($\bet_0$\/) regarding the properties of
$\cS$\/ to two ``Everett copies''
in states of awareness $\bet_1$\/, $\bet_2$\/, i.e., respectively  having seen $\cS$\/ in state $\al_1$\/ and having
seen  seen  $\cS$\/ in state $\al_2$\/.
As in the Schr\"{o}dinger picture ontology, the possible
values of physical properties are those of  the eigenvalues of the corresponding operator; it is necessary to consider only nondegenerate eigenvalues so that
there is no ambiguity regarding the nature and number of the splitting into Everett copies.\footnote{For a more extensive discussion of the ontology of Heisenberg-picture Everett quantum mechanics, see \cite[Sec. 5]{Rubin03}.}

To introduce preclusion, we define the weight associated with  $\wh{b}_i$ \/ to be the matrix element of the
projector\footnote{Termed a 
``label'' in \cite{Rubin01}, since such factors are responsible for ensuring
the correct matching-up of entangled systems and observers in Einstein-Podolsky-Rosen-type scenarios.} $\wh{P}_i^{\cS}$\/ associated with  $\wh{b}_i$\/ between
the initial Heisenberg-picture state and its adjoint:
\be
w_{\;\wh{b},i}=\la\psi_{in}|\wh{P}_i^{\cS} |\psi_{in}\ra.\label{weightHPdef}
\ee
\xxx{weightHPdef}
Then an observer $\cO$\/ with state of awareness $\bet_i$\/ exists at time $t$\/ if 
$w_{\;\wh{b},i} > \eps_P$\/.

The weight  (\ref{weightHPdef}) is of course just the usual value of the probability,  in the state   $|\psi_{in}\ra$\/,  for the outcome
corresponding to the projection operator $ \wh{P}_i^{\cS}$\/ .
We note that even the usual quantum formalism employs hard preclusion with  $\eps_P=0$\/,
since, in any interpretation, components of the state vector with zero norm correspond to outcomes which with certainty do not occur.  
In the 
unphysical limiting case of an infinitely-large ($N=\infty$\/) ensemble of measured systems one
can show that
all outcomes of the measurement of relative frequency inconsistent with the Born rule are precluded even if $\eps_P=0$\/  \cite{Rubin03}.  Preclusion with 
$\eps_P > 0$\/ enables us to implement this approach for the physically far more
interesting case of a finite ensemble.\footnote{Much of the analysis of the finite-ensemble case is similar or identical to the analysis of the infinite-ensemble case presented in \cite{Rubin03}, and the reader is referred to that reference for intermediate steps that have been left out of Sec.~\ref{SecModel}.}

\section{A model of relative-frequency meas\-ure\-ment in \\ Heisenberg-picture Everett quantum mechanics with preclusion}\label{SecModel}
\xxx{SecModel}

Our model for relative-frequency measurement involves three types of quantum systems. Measured
systems $\cS^{(p)}$\/, $p=1,\ldots,N$\/, constitute the ensemble of  identical 
systems required to investigate the relative frequency of a particular outcome. 
The state space of each $\cS^{(p)}$\/ is spanned by the two eigenstates
of the (Schr\"{o}dinger-picture) operator $\wh{a}^{(p)}$\/, 
\be
\wh{a}^{(p)} |\cS^{(p)}; \al_{i^{(p)}} \ra =\al_{i^{(p)}}|\cS^{(p)};\al_{i^{(p)}} \ra,\label{aopaction}
\hs i^{(p)}=1,2, \hs  p=1,\ldots, N,
\ee
\xxx{aopaction}
\be 
\al_{1} \neq  \al_{2}.\label{alphanondegen}
\ee
\xxx{alphanondegen}
For each system $\cS^{(p)}$\/ there is an observer/measuring apparatus $\cO^{(p)}$\/, the state space
of which is spanned by the three eigenvectors of $b^{(p)}$\/,
\be
\wh{b}^{(p)} |\cO^{(p)}; \bet_{i^{(p)}} \ra =\bet_{i^{(p)}}|\cO^{(p)};\bet_{i^{(p)}} \ra,\label{bopaction}
\hs i^{(p)}=0,1,2, \hs  p=1,\ldots, N,
\ee
\xxx{bopaction}
\be 
\bet_{i} \neq  \bet_{j}, \hs i \neq j.\label{betanondegen}
\ee
\xxx{betanondegen}
The  eigenvalues $\bet_0$\/,  $\bet_1$\/ and  $\bet_2$\/ correspond, respectively, to 
 $\cO^{(p)}$\/ being in a state of ignorance, having observed $\al_1$\/, and having observed
$\al_2$\/. Each measurement is implemented by an interaction of the form described in
Sec. \ref{SecEQMHP}.

The third type of system is the relative-frequency observer $\cF$\/, which queries the observers $\cO^{(p)}$\/
and computes the relative frequency for the observation $\bet_1$\/.\footnote{By including the 
$\cO^{(p)}$\/'s in the model, instead of having  $\cF$\/
interact with the $\cS^{(p)}$\/'s directly, we refute claims in
\cite{Stein84,Farhietal89} that measurements of individual systems are incompatible with obtaining the Born rule. See discussion in \cite[Sec. 5]{Rubin03}.}
The possible values of relative frequency, for any fixed value  of $N$\/, are $0$\/,
$1/N$\/, $2/N$\ , $\ldots$\/ \nolinebreak , $(N-1)/N$\/, $1.$\/ As $N$\/ increases without limit  the resolution  needed to distinguish
between adjacent values of relative frequency becomes finer and the total number of possible
values of relative frequency becomes larger. But any given device will only have a limited
resolution \cite{Graham73} as well as a limit on its information-storage capacity set by its construction
but ultimately capped by gravitational considerations  \cite{Beckenstein01}.  Since $\cF$\/ cannot
% neither
%contain storage locations for an arbitrarily large number of
 store an unlimited number of digits, for sufficiently large $N$\/ it  will be unable to record precisely the
value of any possible result for the relative-frequency measurement. Nor, of course, can it retain the 
values obtained from each of the $\cO^{(p)}$\/'s  as $N$\/ increases without limit.

With %these considerations
this in mind, we construct a model of a relative-frequency observer/measuring device $\cF$\/ by
taking the state space of $\cF$\/ to be spanned by 
a discrete finite set of eigenvectors of an operator $\wh{f}$\/:
\be
\wh{f}|\cF;\phi_i\ra =\phi_i |\cF;\phi_i\ra, \hs i=0,\ldots,\nu+1,\label{fopaction}
\ee
\xxx{fopaction}
where
\be
\phi_i= (i-1)/\nu, \hs  i=1,\ldots, \nu+1 \label{phiival}
\ee
\xxx{phiival}
are the possible outputs of the relative-frequency device after the completion of the relative-frequency
measurement, and 
\be
\phi_0=-1/\nu.\label{phi0val}
\ee
\xxx{phi0val}
 corresponds to a state of ignorance. 

Define the relative frequency function for the result $\bet_{i\pup}=\bet_1$\/,
\be
f(\bet_{i^{(1)}}, \ldots, \bet_{i^{(N)}})=(1/N)\sum_{p=1}^N \delta_{i\pup,1},
\hs i\pup \neq 0 \hs \forall \hs p.
\label{rel-freq-func}
\ee
\xxx{rel-freq-func}
The possible values of this function are
\be
%f_k=(k-1)/N, \hs k=1,\ldots,N+1.
f_l=l/N, \hs l=0,\ldots,N. \label{f-l-vals}
\ee
\xxx{f-l-vals}
Define also the finite-resolution relative frequency function
$\phi(\bet_{i^{(1)}}, \ldots, \bet_{i^{(N)}})$\/
to be that $\phi_i$\/, $i=1, \ldots, \nu+1$\/, which is closest in
value to  $f(\bet_{i^{(1)}}, \ldots, \bet_{i^{(N)}})$\/:
%\be
\begin{eqnarray}
\phi(\bet_{i^{(1)}}, \ldots, \bet_{i^{(N)}})&=&
\arg\min_{\phi_i} |\phi_i -f(\bet_{i^{(1)}}, \ldots, \bet_{i^{(N)}})|, \label{fin-res-freq-fun}
\hs i\pup \neq 0 \hs \forall \hs p\\
&& \hbox{\rm (smaller $\phi_i$\/ in case of a tie)} \label{case-of-tie} \nonumber
\end{eqnarray}
\xxx{fin-res-freq-fun}
\xxx{case-of-tie} 
where the $\phi_i$\/'s are as given in (\ref{phiival}).
It will also prove convenient to define
\begin{eqnarray}
\wti{\phi}(\bet_{i^{(1)}}, \ldots, \bet_{i^{(N)}}) &=& \phi(\bet_{i^{(1)}}, \ldots, \bet_{i^{(N)}}), \hs i\pup \neq 0 \hs \forall \hs p\\
&=& \phi_0 \hs \mbox{\rm otherwise} \label{phi-0-otherwise}
\end{eqnarray}
\xxx{phi-0-otherwise}
where $\phi_0$\/ is as given in (\ref{phi0val}).

The unitary operator corresponding to the measurement of all of the $\cO\pup$\/'s
by $\cF$\/ can then be written as
\be
\wh{U}_{\cF}=\sum_{k=0}^{\nu+1}\wh{u}_k^{\cF}\otimes \wh{P}_k^{\wti{\cO}},\label{U-F}
\ee
\xxx{U-F}
where
\be
\wh{u}_k^{\cF}|\cF;\phi_0\ra=|\cF;\phi_k\ra, \hs k=0, \ldots, \nu + 1,
\ee
Note that $\wh{u}_0^{\cF}$\/ acts as the identity on the ignorant state
$|\cF;\phi_0\ra$\/. $\wh{P}_k^{\wti{\cO}}$\/ is the
projection operator which projects out those states corresponding
to $\phi_k$\/:
\be 
\wh{P}_k^{\wti{\cO}}=\sum_{i^{(1)}=0}^2 \ldots \sum_{i^{(N)}=0}^2 
\delta_{\nu \wti{\phi}(\bet_{i^{(1)}}, \ldots, \bet_{i^{(N)}}), k-1}
\otimes_{p=1}^N \wh{P}_{i^{(p)}}^{\cO^{(p)}} \label{P-wticO-def}
\ee
\xxx{P-wticO-def}
where
\be
\wh{P}_{i^{(p)}}^{\cO^{(p)}}=|\cO\pup;\bet_{i\pup}\ra \la \cO\pup;\bet_{i\pup}|.
\label{P-cO-i-proj}
\ee
\xxx{P-cO-i-proj}

%The time-evolution operator which implements the measurement by
%$\cF$\/ of the relative frequency of outcome $\bet_1$\/  in the  ensemble of
%$\cO_{(p)}$\/'s is

At  time $t_1$\/ subsequent to the conclusion of all the measurements---i.e., after each $\cO\pup$\/ measures its corresponding 
$\cS\pup$\/ and all $N$\/ of the $\cO\pup$\/'s are subsequently measured by $\cF$\/---the relative-frequency operator $\wh{f}$\/ has the form

\be
\wh{f}(t_1)=\sum_{k=0}^{\nu + 1} \wh{f}_k \otimes \wh{L}_k,\label{f-k-t1}
\ee
\xxx{f-k-t1}
where
\be
\wh{f}_k=\wh{u}_k^{\cF\dag} \;  \wh{f} \; \wh{u}_k^{\cF}, \hs k=0,\ldots,\nu+1,
\ee
and
\be
\wh{L}_k=\sum_{i^{(1)}=1}^2 \ldots \sum_{i^{(N)}=1}^2 
\left(
\left(\otimes_{p=1}^N \wh{u}_{i^{(p)}}^{(p)\dag}\right) \wh{P}_k^{\wti{\cO}} 
\left(\otimes_{q=1}^N \wh{u}_{i^{(q)}}^{(q)}\right)
\right)
\otimes_{r=1}^N \wh{P}_{i^{(r)}}^{\cS^{(r)}}, \hs k=0,\ldots,\nu+1. \label{L-k-def}
\ee
\xxx{L-k-def}
$\wh{f}(t_1)$\/ in (\ref{f-k-t1}) is of the post-ideal-measurement form (\ref{bt}), so the weight associated with the $k^{\rm th}$\/ possible outcome is
\be
W_{f,k}(t_1)=\la \psi(t_0)| \wh{L}_k | \psi(t_0)\ra.\label{W-fkt-1}
\ee
\xxx {W-fkt-1}
where $| \psi(t_0)\ra$\/ is the constant Heisenberg-picture state. 

We take $| \psi(t_0)\ra$\/ to be the product
state in which $\cF$\/ and the $\cO\pup$\/'s are ignorant and
each of the $\cS\pup$\/'s is in a superposition of $|\cS\pup;\al_1\ra$\/
and $|\cS\pup;\al_2\ra$\/ with the same coefficients:
\be
|\psi(t_0)\ra= |\cF;\phi_0 \ra \prod_{p=1}^N |\cO\pup;\bet_0\ra
				       \prod_{q=1}^N |\cS^{(q)};\psi_0\ra, \label{big-init-state}
\ee
\xxx{big-init-state}
\be
|\cS^{(q)};\psi_0\ra = c_1^{(q)}|\cS\pup;\al_1\ra + c_2^{(q)}|\cS\pup;\al_2\ra,
\hs p=1,\ldots,N, \label{Sp_init_state}
\ee
with
\be
c_1^{(q)}=c_1,\hs c_2^{(q)}=c_2 \hs \forall \; q, \label{c_same}
\ee
\be
|c_1|^2 + |c_2|^2=1. \label{c-sum}
\ee
\xxx{c-sum}

Using (\ref{L-k-def}) and (\ref{big-init-state}) in (\ref{W-fkt-1}), we find (see \cite{Rubin03}, Sec. 3)
\begin{equation}
W_{f,0}(t_1)=0\;\;\forall N.
\end{equation}
%This simply 
That is, at the completion of the relative-frequency measurement there is no Everett copy of \cF\/ which is in a state of ignorance.

The weights corresponding to the other Everett copies of \cF\/ are
\be
%W_{f,k}(t_1)=\sum_{l \; \; | \; \; 0 \leq l \leq N, \;  N(\phi_k-(1/2\nu)) < l \leq N(\phi_k + (1/2\nu))}
%p_{N,l}\label{Wfk}
W_{f,k}(t_1)=\sum_{l \; \; | \; \; 0 \leq l \leq N, \;  N\left(\phi_k-\frac{1}{2\nu}\right) < l \leq N\left(\phi_k + \frac{1}{2\nu}\right)}
p_{N,l}\label{Wfk}
\ee
\xxx{Wfk}
where
\be
%p_{N,l}=\frac{N!}{l!(N-l)!}|c_1|^{2l}|c_2|^{2(N-l)}. \label{p_N_l_def}
p_{N,l}=\frac{N!}{l!(N-l)!}p^{l}q^{(N-l)},\label{p_N_l_def}
\ee
and where we have set
\be
p=|c_1|^2, \hspace*{1cm}q=|c_2|^2,\label{pqdef}
\ee
\xxx{pqdef}
so
\be
q=1-p.\label{qeq1mp}
\ee
\xxx{qeq1mp}

\begin{figure}
\centering
\underline{$| \leftarrow \frac{1}{2\nu} \rightarrow |\leftarrow \frac{1}{2\nu} \rightarrow |
\leftarrow \frac{1}{2\nu} \rightarrow |\leftarrow \frac{1}{2\nu} \rightarrow |$}\\
\vspace*{1mm}
%\vspace*{-.3cm}
%$\hspace*{.74in}\phi_{k\ppr} \hspace*{.66in}%|c_1|^2 
$\hspace*{.74in}\phi_{k\pr-1} \hspace*{.66in}
%p \hspace*{.25in}\phi_{k\pr} \hspace*{.65in}$\\
p \hspace*{.37in}\phi_{k\pr} \hspace*{.65in}$\\$\hspace*{1.405in} | \hspace{-.15cm} \leftarrow \hspace{-.1cm} \Delta \hspace{-.1cm} \rightarrow \hspace{-.15cm}|
\hspace*{.4825in}$\\
\caption{Finite resolution relative frequencies %$\phi_{k\pr-1}$\/, 
%$\phi_{k\pr}$\/ 
and Born-rule  probability $p$\/  % $|c_1|^2 $\/ %in example %of Sec.\ref{subsec_W}
in the case that %$|c_1|^2  
$p < \phi_{k\pr}$\/, $\Delta < \frac{1}{2\nu}$\/.}
\label{Fig1}
\end{figure}

The expression (\ref{Wfk}) for $W_{f,k}(t_1)$\/ can be evaluated by a careful application of the DeMoivre-Laplace theorem. 
Let $\phi_{k\pr}$\/ be that finite-resolution relative frequency which is closest to $p$\/, 
i.e.
\be
k\pr=\arg \min_k |\phi_k-p|,\;\;k=1,\ldots,\nu + 1 \label{kprdef}
\ee 
\xxx{kprdef}
For concreteness suppose that
\be
\phi_{k\pr}> p,\label{phikprgtp}
\ee
\xxx{phikprgtp}
so
\be
\phi_{k\pr}=p+\Delta.\label{Deltadef}
\ee
\xxx{Deltadef}
(See Fig. {\ref{Fig1}.)    
Since the $\phi_{k}$\/'s are uniformly spaced a distance $\frac{1}{\nu}$\/  apart,
\be
0\leq\Delta < \frac{1}{2\nu}.\label{Deltarange}
\ee
\xxx{Deltarange}
We will treat separately the case $\Delta=\frac{1}{2\nu}$\/, i.e., two finite-resolution relative frequencies equally close to $p$\/.

%\textcolor{red}{\bf Correct this one above, see box 2 p.WC-4}\\
Using (\ref{Deltadef}) we can write (\ref{Wfk}), for the case $k=k\pr$\/, as
\be
W_{f,k\pr}(t_1)=\sum_{l \; \; | \; \; 0 \leq l \leq N, \;  
-\sqrt{N}\left(\frac{\frac{1}{2\nu}-\Delta}{\sqrt{pq}}\right)< \frac{l-Np}{\sqrt{Npq}} \leq 
\sqrt{N}\left(\frac{\frac{1}{2\nu}+\Delta}{\sqrt{pq}}\right)}
p_{N,l}\label{Wfkpr}
\ee
\xxx{Wfkpr}
\yyy{bx 1 p. WC-32}
Define
\be
D\pr(A)=
\sum_{l \; \; | \; \; 0 \leq l \leq N, \;  
-A\leq \frac{l-Np}{\sqrt{Npq}} \leq 
A}
p_{N,l}\label{DprAdef}
\ee
\xxx{DprAdef}
\yyy{bx 2 p. WC-32}
where
\be
A>0 \label{Adef}
\ee
is an arbitrary positive number. Provided that $N$\/ is sufficiently large compared to $A$\/, the sum in (\ref{Wfkpr}) will
contain all of the summands in (\ref{DprAdef}) as well as possibly others; and since all summands are positive, the sum in (\ref{Wfkpr}) will then be at least as large as that in (\ref{DprAdef}). Specifically,
\be
W_{f,k\pr}(t_1)_{[N]} \ge D\pr(A)_{[N]}, \hspace*{1cm}\sqrt{N}\left(\frac{\frac{1}{2\nu}\pm\Delta}{\sqrt{pq}}\right) > A.
\label{WfkprgeDprA}
\ee
\xxx{WfkprgeDprA}
\yyy{bxs 1-3 p. WC-50}
(We have added the subscript ``$[N]$\/'' to $W_{f,k\pr}(t_1)_{[N]}$\/ and $D\pr(A)_{[N]}$\/ in (\ref{WfkprgeDprA}) to remind the reader of the dependence of these quantities on $N$\/.) 

Using the  DeMoivre-Laplace theorem \cite[eq. (4.5.2), p. 204]{Renyi70} and the definition of the error function 
\cite[eq. 40:3:1, p. 406]{Oldhametal09},
\be 
\lim_{N \rightarrow \infty} D\pr(A)_{[N]}=\erf(A/\sqrt{2}),\label{DprANerfA}
\ee
\xxx{DprANerfA}
\yyy{bx 4 p. WC-50}
This implies that, for any positive number $\eps_D$\/, there exists an integer $N_D(\eps_D,A)$\/ such that
\be
D\pr(A)_{[N]} > \erf(A/\sqrt{2})-\eps_D,\hspace*{1cm} N > N_D(\eps_D,A)\label{DprAepslim}
\ee
\xxx{DprAepslim}
\yyy{bx 5 p. WC-50}
From (\ref{WfkprgeDprA}) and (\ref{DprAepslim}) we obtain
\be
W_{f,k\pr}(t_1)_{[N]} > \erf(A/\sqrt{2}) - \eps_D,\hspace*{.75cm} N> \max\left(N_D(\eps_D,A),\frac{A^2pq}{\left(\frac{1}{2\nu}-\Delta\right)^2}\right)\!\!, \hspace*{.325cm} \eps_D > 0 \label{Wfkprgterf}
\ee
\xxx{Wfkprgterf}
\yyy{bx 6 p. WC-50} 
From the definition of the error function it follows that
\be
\lim_{A \rightarrow \infty}\erf(A/\sqrt{2})=1.\label{inflimerf}
\ee
\xxx{inflimerf}
So for any positive number $\eps_e$\/ there exists a number  $A_e(\eps_e)$\/ such that
\be
\erf(A/\sqrt{2}) > 1-\eps_e,\hspace*{1cm} A> A_e(\eps_e).\label{lolimerf}
\ee
\xxx{lolimerf}
\yyy{bx 8 p.WC-50}
Using (\ref{lolimerf}) in (\ref{Wfkprgterf}), we obtain
$$
W_{f,k\pr}(t_1)_{[N]} > 1 -\eps_e- \eps_D, \hspace*{1cm} N> \max\left(N_D(\eps_D,A),\frac{A^2pq}{\left(\frac{1}{2\nu}-\Delta\right)^2}\right)\!\!, \hspace*{2mm} A> A_e(\eps_e),
$$
\be
\eps_D, \eps_e > 0 
\label{Wfklteps}
\ee
\xxx{Wfklteps}
\yyy{last bx p. WC-50}
Let $\eps$\/ be a number satsifying
\be
\eps > \eps_D+ \eps_e. \label{epsdef}
\ee
\xxx{epsdef}
\yyy{bx 1 p WC-63}
Then
$$
W_{f,k\pr}(t_1)_{[N]} > 1 -\eps, \hspace*{2mm} N> \max\left(N_D(\eps_D,A),\frac{A^2pq}{\left(\frac{1}{2\nu}-\Delta\right)^2}\right)\!\!, \hspace*{2mm} A> A_e(\eps_e), \hspace*{2mm}\eps > \eps_D+\eps_e, \nonumber
$$
\be
\eps_D, \eps_e, \eps > 0 \label{Wfklteps2}
\ee
\xxx{Wfklteps2}
\yyy{bx 5 p. WC-63}

So $W_{f,k\pr}(t_1)$\/, the weight for the finite-resolution relative frequency closest to the Born-rule relative frequency $p$\/, can be brought arbitrarily close to unity  by making the size $N$\/ of the spin ensemble being measured sufficiently large.\footnote{The conditions in (\ref
{Wfklteps2}) are to be read from right to left: Pick a desired positive $\eps$\/, then choose a positive  $\eps_D$\/  and a positive   $\eps_e$\/ such that their sum is less than $\eps$\/.  Then choose a positive number $A$\/ greater than $A_e(\eps_e)$\/.  Using $A$\/ and $\eps_D$\/ we can now select a value for $N$\/ large enough so that $W_{f,k\pr}(t_1)$\/ is within the desired distance $\eps$\/ of 1.}

Of course $W_{f,k\pr}(t_1)$\/ will never be {\em larger}\/  than unity. The $W_{f,k}$\/'s sum to unity \cite[eq. (100)]{Rubin03},%{\bf (CITE RF PAPER)}
\be
\sum_{k=1}^{\nu+1}W_{f,k}(t_1)=1, \label{Wsumunity}
\ee
\xxx{Wsumunity}
and are all positive,
\be
W_{f,k}(t_1) > 0, \hspace*{1cm} k=1,\ldots,\nu+1\label{Wfkpos}
\ee
\xxx{Wfkpos}
so
\be
W_{f,k\pr}(t_1) < 1.\label{Wfklt1}
\ee
\xxx{Wfklt1}
\yyy{p WC-64}

Using (\ref{Wfklteps2}) with (\ref{Wsumunity}) with (\ref{Wfkpos}) we obtain
$$ 
W_{f,k}(t_1) < \eps, \hspace*{.75cm} k \neq k\pr,\hspace*{2mm} N> \max\left(N_D(\eps_D,A),\frac{A^2pq}{\left(\frac{1}{2\nu}-\Delta\right)^2}\right), \hspace*{2mm} A> A_e(\eps_e), \hspace*{2mm}\eps > \eps_D+\eps_e,
$$
\be
\eps_D, \eps_e, \eps > 0 \label{Wfklteps3}
\ee
\xxx{Wfklteps3}
\yyy{last bx p. WC-63}

I.e., the same $N$\/ that yields $W_{f,k\pr}(t_1)$\/ within $\eps$\/ of $1$\/ will ensure that any other $W_{f,k}(t_1)$\/, $k\neq k\pr$\/, will be within $\eps$\/ of 0. For any preclusion threshold $\eps_P$\/ between $\eps$\/ and $1-\eps$\/, Everett copies of the relative-frequency observer $\cF$\/ perceiving
the non-Born finite-resolution relative frequencies $\phi_k$\/,  $k\neq k\pr$\/, will be precluded while the Everett copy perceiving the finite-resolution relative frequency $\phi_{k\pr}$\/ closest to the Born value will survive. This conclusion also holds if the closest finite-resolution relative frequency to $p$\/ is less than $p$\/.

What about the case $\Delta=\frac{1}{2\nu}$\/, i.e., $p$\/  equidistant from the two closest finite-resolution relative frequencies?  Denote these $\phi_k$\/'s by
\be
\phi_{k_<}=p-\frac{1}{2\nu}, \label{phiklt}
\ee
\xxx{phiklt}
\yyy{bx 1 p WC-20}
\be
\phi_{k_>}=p+\frac{1}{2\nu}.  \label{phikgt}
\ee 
\xxx{phikgt}
\yyy{bx 2 p WC-20} 
Using (\ref{Wfk}) and (\ref{phiklt}),
\be
W_{f,k_<}(t_1)=\sum_{l \; \; | \; \; 0 \leq l \leq N, \;  
-\frac{1}{\nu}\sqrt{\frac{N}{pq}}< \frac{l-Np}{\sqrt{Npq}} \leq 
0}
p_{N,l}.\label{Wfklt}
\ee
\xxx{Wfklt}
\yyy{bx 1 p. WC-21}
This can be evaluated in essentially the same manner as $W_{f,k\pr}(t_1)$\/ above.

Define
\be
D_<(A)=
\sum_{l \; \; | \; \; 0 \leq l \leq N, \;  
-A\leq \frac{l-Np}{\sqrt{Npq}} \leq 
0}
p_{N,l},\hspace*{1cm}A>0\label{DltAdef}
\ee
\xxx{DltAdef}
\yyy{bx 2 p. WC-21}
Then
\be
W_{f,k_<}(t_1) \ge D_<(A),\hspace*{1cm}\frac{1}{\nu}\sqrt{\frac{N}{pq}} > A \label{WfkltgteDltA}
\ee
\xxx{WfkltgteDltA}
\yyy{bx 4 p WC-21}
\xyz{\textcolor{green}{Mon. Oct 26, 2020 12:37PM}}\\
The DeMoivre-Laplace theorem implies that, for any positive number $\eps_D$\/, there exists a number $N_<(\eps_D,A)$\/ such that
\be 
D_<(A) > \frac{1}{2}\erf(A/\sqrt{2}) - \eps_D, \hspace*{1cm} N > N_<(\eps_D,A) \label{DltAlim}
\ee
\xxx{DltAlim}
\yyy{bx 3 p WC-54}
Using this and (\ref{lolimerf}) in (\ref{WfkltgteDltA})
\be
W_{f,k_<}(t_1) > \frac{1}{2}-\frac{\eps_e}{2}-\eps_D,\hspace*{1cm}
N > \max \left(N_<(\eps_D,A), \nu^2 A^2 pq\right)\!\!,\hspace*{2mm}A > A_e(\eps_e),\hspace*{2mm}\eps_D,\eps_e > 0.
\label{Wfkltgt}
\ee
\xxx{Wfkltgt}
\yyy{bx 2 p WC-55}
Let $\eps_<$\/ be a positive number satisfying
\be
\eps_< > \frac{\eps_e}{2}+\eps_D. \label{epslt}
\ee
\xxx{epslt}
\yyy{bx 1,3 p. WC-60}
\xyz{\textcolor{green}{Mon. Oct 26, 2020 1:15PM}}\\
\xyz{\textcolor{green}{Mon. Oct 26, 2020 2:15PM}}\\
Then 
$$
W_{f,k_<}(t_1) > \frac{1}{2}-\eps,\hspace*{1cm}
N > \max \left(N_<(\eps_D,A), \nu^2 A^2 pq\right)\!\!,\hspace*{2mm}A > A_e(\eps_e),\hspace*{2mm}\eps_< > \frac{\eps_e}{2}+\eps_D,
$$
\be
\eps_D, \eps_e, \eps_< > 0.
\label{Wfkltgt2}
\ee
\xxx{Wfkltgt2}
\yyy{bx 8 p WC-60}

Using (\ref{phikgt}) in (\ref{Wfk}),
\be
W_{f,k_>}(t_1)=\sum_{l \;\;|\;\; 0 \leq l \leq N,\; 0< \frac{l-Np}{\sqrt{Npq}} \leq \frac{1}{\nu}\sqrt{\frac{N}{pq}}}p_{N,l}.\label{Wfkgt}
\ee
\xxx{Wfkgt}
\yyy{bx 2 p. WC-23}
To evaluate (\ref{Wfkgt}), define
\be
\wti{W}_{f,k_>}(t_1)=\sum_{l \;\;|\;\; 0 \leq l \leq N,\; 0 \leq \frac{l-Np}{\sqrt{Npq}} \leq \frac{1}{\nu}\sqrt{\frac{N}{pq}}}p_{N,l}\label{Wfktigt}
\ee
\xxx{Wfktigt}
\yyy{bx 1 p. WC-24}
and
\be
D_>(A)=
\sum_{l \; \; | \; \; 0 \leq l \leq N, \;  
0 \leq \frac{l-Np}{\sqrt{Npq}} \leq 
A}
p_{N,l},\hspace*{1cm}A>0.\label{DgtAdef}
\ee
\xxx{DgtAdef}
\yyy{bx 2 p. WC-24}
These satisfy
\be
\wti{W}_{f,k_>}(t_1) \ge D_>(A), \hspace*{1cm}\frac{1}{\nu}\sqrt{\frac{N}{pq}} > A. \label{WfktigtgtDgtA}
\ee
\xxx{WfktigtgtDgtA}
\yyy{bx3 p WC-55}
\xyz{\textcolor{green}{Tues. Oct 27, 2020 8:43AM}}\\
Using (\ref{lolimerf}) and (\ref{DltAlim}) in (\ref{WfktigtgtDgtA}),
\be
\wti{W}_{f,k_>}(t_1) > \frac{1}{2}-\frac{\eps_e}{2}-\eps_D,\hspace*{1cm}
N > \max \left(N_<(\eps_D,A), \nu^2 A^2 pq\right)\!\!,\hspace*{2mm}A > A_e(\eps_e),\hspace*{2mm}\eps_D,\eps_e > 0.
\label{Wfkgtgt}
\ee
\xxx{Wfkgtgt}
\yyy{bx 3 p WC-56}
(Note that the conditions on $N$\/ are the same as for  $W_{f,k_<}(t_1)$\/ in (\ref{Wfkltgt}).)
\\ \xyz{\textcolor{green}{Tues. Oct 27, 2020 9:05AM}}\\
\xyz{\textcolor{green}{Tues. Oct 27, 2020 9:51AM}}\\

From (\ref{Wfkgt}) and (\ref{Wfktigt}),
\be
\begin{array}{cccr}
\wti{W}_{f,k_>}-W_{f,k_>} &=& 0,                 & \lfloor Np \rfloor \neq Np, \\
                                         &=& p_{N,[Np]},   & \lfloor Np \rfloor = Np, \label{WfktiminusWfk}
\end{array}
\ee
\xxx{WfktiminusWfk}
\yyy{bxs 4,5 p WC-56}
where $\lfloor x \rfloor $\/ is the floor of $x$\/.  Therefore
\be
\wti{W}_{f,k_>}-W_{f,k_>}  \leq p_{N,[Np]}.  \label{WtiWltpNNp}
\ee
\xxx{WtiWltpNNp}
\yyy{bx 6 p. WC-56}
Using (\ref{Wfkgtgt})   and (\ref{WtiWltpNNp}),
$$
W_{f,k_>}(t_1) > \frac{1}{2}-\frac{\eps_e}{2}-\eps_D-p_{N,\lfloor Np \rfloor},\hspace*{1cm}
N > \max \left(N_<(\eps_D,A), \nu^2 A^2 pq\right)\!\!,\hspace*{2mm}A > A_e(\eps_e),
$$
\be
\eps_D,\eps_e > 0.
\label{Wfkgtgt2}
\ee
\xxx{Wfkgtgt2}
\yyy{bx 1 p WC-57} 
The DeMoivre-Laplace theorem\cite[eq. (4.5.1), p. 204]{Renyi70} gives an asymptotic expression for $p_{N,l}$\/,
\be
p_{N,l} \sim \frac{e^{-(l-Np)^2/2Npq}}{\sqrt{2\pi Npq}},\hspace*{1cm} |l-Np|=o(N^{2/3}).\label{pNlasymptotic}
\ee
\xxx{pNlasymptotic}
\yyy{last bx p WC-69}
Since
\be
| \lfloor Np \rfloor - Np | < 1,\label{Npfloordiflim}
\ee
\xxx{Npfloordiflim}
we can make use of (\ref{pNlasymptotic}) with $l=\lfloor Np \rfloor$\/ to conclude that, for any positive number $\eps_\ast$\/, there is an integer $N_\ast(\eps_\ast,p)$\/ such that 
\be
p_{N,\lfloor Np\rfloor }< \frac{1+ \eps_\ast}{\sqrt{2 \pi N p q}},\hspace*{1cm}N > N_\ast(\eps_\ast,p).\label{pNNplim}
\ee
\xxx{pNNplim}
\yyy{bx 3 p. WC-72}
(The limit involved in (\ref{pNlasymptotic}) is uniform in $l$\/ \cite[footnote p.~204]{Renyi70}, so $N_\ast(\eps_\ast,p)$\/ does not depend on $l$\/.) From (\ref{Wfkgtgt2}) and (\ref{pNNplim}) we obtain
$$
W_{f,k_>}(t_1) > \frac{1}{2}-\frac{\eps_e}{2}-\eps_D- \frac{1+ \eps_\ast}{\sqrt{2 \pi N p q}} ,\hspace*{1cm}
N > \max \left(N_<(\eps_D,A), \nu^2 A^2 pq, N_\ast(\eps_\ast,p)\right)\!\!,
$$
\be
A > A_e(\eps_e),\hspace*{2mm} \eps_D,\eps_e, \eps_\ast > 0.
\label{Wfkgtgt3}
\ee
\xxx{Wfkgtgt3}
\yyy{bx 1 p WC-58} 
\\ \xyz{\textcolor{green}{Thurs. Oct 29, 2020 11:31AM}}\\
Let $\eps_>$\/ be a positive number satisfying
\be
\eps_> > \frac{\eps_e}{2}+\eps_D+ \frac{1+ \eps_\ast}{\sqrt{2 \pi N p q}}. \label{epsgtdef}
\ee
\xxx{epsgtdef}
\yyy{top eq p WC-59}
Since $\eps_\ast$\/ is positive, this implies
\be
\eps_> > \frac{\eps_e}{2}+\eps_D \label{epsgtepss}
\ee
\xxx{epsgtepss}
\yyy{based on stuff on p WC-59}
which, with (\ref{epsgtdef}), yields
\be
\sqrt{N}> \frac{1+ \eps_\ast}{\left(\eps_>- \frac{\eps_e}{2}-\eps_D \right)\sqrt{2 \pi p q}}. \label{sqrtNlim}
\ee
\xxx{sqrtNlim}
\yyy{from bx 1 p. WC-59}
Conversely, (\ref{epsgtepss}) and (\ref{sqrtNlim}) give (\ref{epsgtdef}).  Using (\ref{Wfkgtgt3})-(\ref{sqrtNlim}) we obtain
$$
W_{f,k_>}(t_1) > \frac{1}{2}-\eps_>,\hspace*{1cm}
N > \max \left(N_<(\eps_D,A), \nu^2 A^2 pq, N_\ast(\eps_\ast,p),
\frac{\left(1+ \eps_\ast\right)^2}{\left(\eps_>- \frac{\eps_e}{2}-\eps_D \right)^2 2 \pi p q}\right)\!\!,
$$
\be
A > A_e(\eps_e),\hspace*{2mm}\eps_> > \frac{\eps_e}{2}+\eps_D,\hspace*{2mm} \eps_D,\eps_e, \eps_\ast,\eps_> > 0.
\label{Wfkgtgt4}
\ee
\xxx{Wfkgtgt4}
\yyy{bx 3 p WC-59}  
Setting
\be
\eps_< =\eps_>=\eps, \label{equalepss}
\ee
\xxx{equalepss}
\yyy{bx 1 p WC-61}
 (\ref{Wfkltgt2}) and (\ref{Wfkgtgt4})  give
$$
W_{f,k_<}(t_1) > \frac{1}{2}-\eps \hspace*{2mm} \mbox{\rm and} \hspace*{2mm} W_{f,k_>}(t_1) > \frac{1}{2}-\eps,
$$
$$
N > \max \left(N_<(\eps_D,A), \nu^2 A^2 pq, N_\ast(\eps_\ast,p),
\frac{\left(1+ \eps_\ast\right)^2}{\left(\eps_>- \frac{\eps_e}{2}-\eps_D \right)^2 2 \pi p q}\right)\!\!,
$$
\be
A > A_e(\eps_e),\hspace*{2mm}\eps_> > \frac{\eps_e}{2}+\eps_D,\hspace*{2mm} \eps_D,\eps_e, \eps_\ast,\eps > 0.
\label{Wfkbothgt4}
\ee
\xxx{Wfkbothgt4}
\yyy{bx 2 p WC-61}  
\xyz{\textcolor{green}{Thurs. Oct 29, 2020 ~3:30PM}}\\
\xyz{\textcolor{green}{Fri. Oct 30, 2020 8:54AM}}\\
Using (\ref{Wsumunity}) and (\ref{Wfkpos}) with (\ref{Wfkbothgt4}) we obtain
$$
W_{f,k}(t_1) < 2\eps,  \hspace*{2mm}k \neq k_<,  k_>,
$$
$$
N > \max \left(N_<(\eps_D,A), \nu^2 A^2 pq, N_\ast(\eps_\ast,p),
\frac{\left(1+ \eps_\ast\right)^2}{\left(\eps_>- \frac{\eps_e}{2}-\eps_D \right)^2 2 \pi p q}\right)\!\!,
$$
\be
A > A_e(\eps_e),\hspace*{2mm}\eps_> > \frac{\eps_e}{2}+\eps_D,\hspace*{2mm} \eps_D,\eps_e, \eps_\ast,\eps > 0.
\label{Wfkother}
\ee
\xxx{Wfkother}
\yyy{last bx p WC-61}

So, if the two  finite-resolution relative frequencies $\phi_{k_<}$\/ and $\phi_{k_>}$\/ closest to $p$\/ are equidistant from $p$\/,
the same $N$\/ that yields $W_{f,k_<}(t_1)$\/ and $W_{f,k_>}(t_1)$\/ within $\eps$\/ of $1/2$\/ will ensure that any other $W_{f,k}(t_1)$\/, $k\neq k_<,k_>$\/, will be within $2\eps$\/ of 0. For any preclusion threshold $\eps_P$\/ between $2\eps$\/ and $1/2-\eps$\/,  Everett copies of the relative-frequency observer $\cF$\/ perceiving
   the non-Born finite-resolution relative frequencies $\phi_k$\/,  $k\neq k_<, k_>$\/, will be precluded while  the two Everett copies perceiving the finite-resolution  relative frequencies $\phi_{k_<}$\/,  $\phi_{k_>}$\/ closest to the Born value will survive.\footnote{By considering the $N \rightarrow \infty$\/ limit of (\ref{Wfkbothgt4}) and (\ref{Wfkother}) we extend the results of \cite[Sec. 3.5]{Rubin03} for $\Delta=1/2\nu$\/ to include the cases in which   $\nu$\/  is even or $p$\/ differs from $1/2$\/.\label{footnotenumber}} \label{footnotepage}

\section{Precluded contingent events}\label{SecPreclusionConsCon}
\xxx{SecPreclusionConsCon}

Any theory of probability, classical or quantum, that employs hard preclusion must deal with apparently-paradoxical situations of the following sort.  
Consider a probabilistic event the occurrence of which is contingent on the occurrence of at least one of a set of $N$\/ antecedent probabilistic events.  Say the probability of any one of the antecedent events is $p$\/.  Then if $p \ll 1$\/, the probability of the contingent event is approximately $Np$\/. If the preclusion threshold $\eps_P$\/ is such that $p < \eps_P < Np$\/, the contingent event will occur despite none of the antecedent events required for its occurrence having occurred.\footnote{I thank an anonymous referee of \cite{Rubin10}  for pointing out this issue to me.}

To avoid such situations we modify the preclusion formalism we have been employing so far by making the preclusion threshold $\eps_P$\/ a function of the dimension $D$\/ of the projection operator corresponding to the event which is (possibly) being precluded.  Specifically, 
\be
\eps_P=\eps_P(D)=D\eps_1 \label{eps1def}
\ee
\xxx{eps1def}
where $\eps_1$\/ is the  preclusion threshold for 1-dimensional projectors.   Note that in the Schr\"{o}dinger picture this dimension-dependence of the preclusion threshold follows automatically from the formalism, in which events are precluded based on the norms of (product) state vectors to which they correspond; see BHZ. 

As an illustration, consider an ``at-least-one'' observer \cG\/ that measures whether one or more of a set of qubits  
in the state $|\Psi\ra$\/ of (\ref{productstate}) %$|\cS^{(1)};\al_{i^{(1)}}\ra$\/,
%$|\cS^{(2)};\al_{i^{(2)}}\ra$\/, \ldots $|\cS^{(N)};\al_{i^{(N)}}\ra$\/ 
is in state $\al_1$\/. The two states of the observer correspond, respectively,  to ``all qubits in state 1'' and ``at least one qubit in state 2:''
\bea
\wh{g}|\cG;\ga_0\ra&=&\ga_0|\cG;\ga_0\ra\hspace*{.5cm}\mbox{\rm (all qubits in state 1)},\\ \label{gamma0def}
\wh{g}|\cG;\ga_1\ra&=&\ga_1|\cG;\ga_1\ra\hspace*{.5cm}\mbox{\rm (at least one qubit in state 2).} \label{gamma1def}
\eea
\xxx{gamma0def}
\xxx{gamma1def}
\yyy{bx. 2 p. P3-31}
For simplicity we do not include observers of the individual qubits 
or a separate ready state for \cG\/, the role of the latter being played by $|\cG;\ga_0\ra$\/. I.e., \cG\/ is  in  state $|\cG;\ga_0\ra$\/ before the measurement, at time $t_{in}$\/, and and remains in that state after completion of the measurement at time $t$\/ if all of the qubits are in state 1, else transitions to state $|\cG;\ga_1\ra$\/.

The unitary transformation which effects this dynamics is
\be
\wh{U}^{\cG\cS}(t)=\exp\left(-i\wh{H}^{\cG\cS}(t-t_{in})\right) \label{UGSdef}
\ee
\xxx{UGSdef}
\yyy{bx. p. P3-31}
where 
\be
\wh{H}^{\cG\cS}=\wh{h}^\cG \otimes \wh{P}_\cG^\cS  \label{HGSdef}
\ee
\xxx{HGSdef}
\yyy{p. 4, p. P3-31}
with
\be
\wh{h}^{\cG}=i\kappa \left(|\cG;\ga_1\ra\la\cG;\ga_0|-|\cG;\ga_0\ra\la\cG;\ga_1|\right) \label{hGdef}
\ee 
\xxx{hGdef}
\yyy{bx. 1, p. P3-32}
and, using (\ref{PSdef}),
\be
\wh{P}_\cG^S=1-\prod_{p=1}^N \wh{P}_1^{\cS^{(p)}}. \label{PGSdef}
\ee
\xxx{PGSdef}
\yyy{bx. 2, p. ALO-1}
Using (\ref{gamma0def})-(\ref{PGSdef}), we find the unitary transformation, at time $t$\/, to be
\be
\wh{U}^{\cG\cS}(t)=\wh{P}^\cS_{\overline{ \cG}}+ \left[\cos(\ka(t-t_{in})+i \sin(\ka(t-t_{in}))\frac{\wh{h}^\cG}{\ka}\right]\otimes \wh{P}^\cS_\cG, \label{UGSt} 
\ee
\xxx{UGSt}
\yyy{P3-38, bx1}
where
\be
\wh{P}^\cS_{\overline{ \cG}}=1-\wh{P}^\cS_\cG=\prod_{p=1}^N \wh{P}_1^{\cS^{(p)}}  \label{PGbarSdef}
\ee
\xxx{PGbarSdef}
\yyy{last bx. p. P3-37 (or is that PS-37?), 3d-to-last bx. p. P3-31}
The Heisenberg-picture at-least-one operator at time $t$\/ is then
\bea
\wh{g}(t)&=&\wh{U}^{\cG\cS\dag}\wh{g}\wh{U}^{\cG\cS\dag}\\ \nonumber
          &=&\wh{g}\otimes\wh{P}^\cS_{\overline{ \cG}}+\left[\cos^2(\ka(t-t_{in}))\wh{g} 
                  +i\sin(\ka(t-t_{in}))\cos(\ka(t-t_{in}))\frac{1}{\ka}[\wh{g},\wh{h}^\cG]  \right. \\ \nonumber
                 &&\hspace*{.75in}\left. +\sin^2(\ka(t-t_{in})\wh{g}_1 \right] \otimes\wh{P}^\cS_\cG \label{gt}
\eea
\xxx{gt}
\yyy{P3-41 last bx, bx. 4 p. P3-43}}
where
\bea
\wh{g}_1&=&\frac{1}{\ka^2}\wh{h}^\cG \wh{g}\wh{h}^\cG \\ \nonumber
&=&\ga_1|\cG;\ga_0\ra\la\cG;\ga_0|+\ga_0|\cG;\ga_1\ra\la\cG;\ga_1|.\label{g1def}
\eea
\xxx{g1def}
\yyy{P3-43, bx 4,5}
Choosing the time $t$\/ to be the time $t_1$\/ at which the measurement is complete (see eq. (\ref{tval})),
the Heisenberg-picture at-least-one operator is in the post-ideal-measurement form,
\be
\wh{g}(t_1)=\wh{g}\otimes\wh{P}^\cS_{\overline{ \cG}} +\wh{g}_1\otimes \wh{P}^\cS_\cG\label{gt1}
\ee
\xxx{gt1}
\yyy{P3-43, last bx}
The weight for $\cG$\/ having determined that at least one of the systems $\cS^{(q)}$\/ is in state 2 is
\be
W^\cS_1=\la \Psi| \wh{P}^\cS_\cG  |\Psi\ra.\label{WS1}
\ee
\xxx{WS1}
\yyy{bx. 1, p. P3-45}
%where $|\psi^\cS\ra$\/ is as in (\ref{productstate}).  
Using (\ref{productstate}),  (\ref{normalization}), (\ref{pqdef})  and (\ref{PGSdef}),
\be
W^\cS_1=1-(1-q)^N.\label{WS1b}
%&\approx &Nq, \hspace*{.5 cm}Nq \ll 1 
\ee
\xxx{WS1b}
\yyy{bx. 2, p. ALO-4}

Suppose now that the probability $q$\/ of any individual qubit being in the state 2 is small enough to be precluded, i.e.,
\be
q < \eps_1\label{qubitprecluded}
\ee
\xxx{qubitprecluded}
\yyy{pp. PCE 1-2}
so of course
\be
Nq < N\eps_1\label{qubitprecluded2}
\ee
\yyy{pp. PCE 1-2}
But 
\be
1-(1-q)^N \le Nq
\ee
\yyy{pp. PCE 1-2}
(induction on $N$\/). So, whenever the outcome of any qubit being in state 2 is precluded, the rule  (\ref{eps1def}), with the dimension $D$\/ equal to $N$\/, ensures that the outcome of $\cG$\/ having detected at least one of the qubits in state 2 is also precluded:
\be
W^\cS_1 < \eps_P(N).\label{WS1precluded}
\ee
\xxx{WS1precluded}
\yyy{pp. PCE 1-2}

\section{Ordinary probabilistic events and biological evolution}\label{SecOrdProbEventsBiolEvol}
\xxx{SecOrdProbEventsBiolEvol}

\subsection{The connection between extraordinary and ordinary probabilistic events}\label{SecConnection}
\xxx{SecConnection}

Accounting for the nonobservation of extraordinary probabilistic events is no small thing; the laws of thermodynamics play a key role in making our world what it is.  But of course there are %a 
myriad %of  
ordinary  probabilistic phenomena encountered in physics and in day-to-day life. How, if at all, are these related to extraordinary probabilistic events and preclusion? \xyz{\textcolor{green}{\bf Fri Nov. 29 8:08 PM}}

There is no shortage of events of low probability which are known to have occurred.  Therefore the preclusion parameter $\eps_P(N)$\/ for sub-Avogadro values of $N$\/ commonly encountered in physics and in life must be extremely small\footnote{Estimates for the size of the preclusion parameter, based on quantum-gravitational  considerations, are presented in BHZ.  Note that the parameter $\eps$\/ of  BHZ is related to the parameter $\eps_1$\/ of the present paper by $\eps=\sqrt{\eps_1}$\/.} in order not to be in conflict with experience, much smaller than the probability of a deviation of the empirical relative frequency from the theoretical probability.  Preclusion, it might seem,  can have no more direct connection with ordinary probabilistic phenomena than can the limiting case of an infinitely large system. 

Again it makes sense to appeal directly to the empirically-observed features associated with the phenomena we are trying to understand, in this case ordinary probabilistic  phenomena as defined in Sec.~\ref{SecIntro}. \xyz{\textcolor{green}{\bf Fri Nov. 29 9:30:04 PM}} The  {\em objective }\/ features associated with ordinary probabilistic phenomena are precisely the {\em subjective}\/ judgements we have regarding them. In the words of de Finetti \cite[p. 174]{deFinetti89}, ``What do we mean when we say, in ordinary language, that an event is more or less probable? We mean that we would be more or less surprised to learn that it has not happened.'' These judgements, also termed ``subjective probabilities'' or ``credences,''  are subjective in that they are the beliefs of individuals; but their existence is an objective fact. 

A feature of subjective judgements of probability  is that, in many though by no means in all cases, they are more or less in accord with the quantities given by physical theories regarding ordinary probabilistic phenomena which we usually think of as ``objective probabilities.''  This connection between subjective judgement and objective-probability-as-given-by-physical-theory bears a resemblance to the Principal Principle of Lewis \cite{Lewis80}  \xyz{\textcolor{red}{\bf START HERE Fri., Nov. 6, 2:47AM}} which, as formulated by 
Papineau \cite{Papineau10}, states that ``it it rational to set your subjective probabilities equal to the objective probabilities.''  In the present context, where  no notion of objective probability governing ordinary probabilistic phenomena has been introduced, what we wish to explain is what we might call the ``empirical Principal Principal,'' the  observation that in many cases subjective judgements assigned to ordinary probabilistic phenomena are consistent with the predictions of physical theory, i.e., with the probabilities assigned by the Born rule based on the same quantum-mechanical amplitudes that, for Avogadro-number-sized ensembles, would lead to the preclusion of extraordinary probabilistic phenomena.\footnote{The status of the Principal Principle in decision-theoretic approaches to probability in Everett quantum mechanics is discussed in \cite{BrownBenPorath20}, and a  derivation thereof in the context of non-Everett quantum mechanics is given in \cite{Earman18}.  }

We have argued above that the preclusion of extraordinary probabilistic phenomena cannot be directly related to the kind of probability witnessed in ordinary probabilistic phenomena essentially because the probabilities involved in the latter are too large.  However, if the phenomenon of subjective judgement of probability is one that comes about through {\em  biological evolution}\/, then the requisite small quantum weights  can arise not through the growth in the size of an ensemble at a single time, but rather through the repetition of interactions between organisms and their environment over  generations upon generations of geological time.  After a sufficiently large number of generations, the existence of species with subjective judgements of probability that are suited for survival in environments that have lower quantum-mechanical amplitude will be precluded.
Surviving species will be those with subjective judgements of probability suited for survival in environments with higher quantum-mechanical amplitude.  The latter, it can be argued, will be environments in which frequencies are close to those required by the Born rule, so subjective judgements of probability will match Born-rule values, the objective probabilities of standard quantum theory.  In the next two subsections we present a model illustrating this process.\footnote{The idea that preclusion explains evolution which then explains subjective probability was presented in my previous paper \cite{Rubin10}, though not the supporting model of Secs. \ref{biolevolhardprecl} and  \ref{subsecBornlikemaverickstates}  below.}

\subsection{Biological evolution from hard preclusion} \label{biolevolhardprecl}
\xxx{biolevolhardprecl}

Existing models of biological evolution describe ordinary probabilistic phenomena, including evolutionary change from one generation to the next, in terms of objective probability; e.g., a species of predator with slightly superior eyesight will in each generation likely do somewhat better at surviving than a similar species with poorer eyesight, and over many generations will with high probability supplant the less sharp-sighted species.   For us to argue that subjective probability, or for that matter any other observed phenomenon, is the result of biological evolution, we require a model of biological evolution which does not require for its description the same ordinary probabilistic phenomena the origin of which we wish to explain, but which rather makes use only of the physical concepts we have at our disposal, namely unitary quantum-mechanical time evolution and hard preclusion.   

Here we present a basic proof-of-concept version of such a model.  For simplicity we work in the 
Schr\"{o}dinger picture with discrete time-steps, one per generation,  and do not model competition between species but only natural selection.  
%What we want to demonstrate is that, after a sufficient number of generations, the quantum-mechanical weight for a species which is a poorer fit to the %environment to survive will drop below the preclusion threshold, while  that for a better-fit species will remain above that threshold.  
Specifically, the model demonstrates that the weight for the survival of a species which is fit to a lower-weight (less ``probable'') environment eventually drops below the preclusion threshold while that for a species fit to a higher-weight (more ``probable'') environment remains above that threshold.

An issue that has to be kept in mind in constructing a model of this sort is the tension between the unitarity  of quantum mechanics and the  irreversibility of the notion of ``extinction.'' 
This is of course not an insuperable obstacle. Following the approach of other models of irreversible processes in quantum mechanics (see, e.g., 
the discussion of the decay of an excited state in \cite[Sec. 6.1]{Davies74}), we set up a unitary framework in which the transition of a species from an extinct state to an alive state is allowed, then  apply the  framework to a time regime small compared to the time required for a return from the extinct state to the alive state.

The objects in the model are an environment {\cE} and two species of organisms ${\cO_B}$\/ and ${\cO_M}$\/.  The environment is represented by a tensor product of states labeled by a discrete time parameter $t$\/: 
\begin{equation}
|\cE\ra=\prod_{t=0}^T |\cE_t\ra. \label{envprod}
\end{equation}
\xxx{envprod}
\yyy{Inferred from bxs 1,2 p HQ82 (p 49 is missing), see also bx 5 p HQ129}
The state $|\cE_t\ra$\/ represents the state of that environment that the organisms interact with at time $t$\/. 
I.e., each $|\cE_t\ra$\/, $t=0,\ldots,T$\/, is the state vector of a distinct entity, as indicated in (\ref{envprod}).  $|\cE_t\ra$\/ is unchanged under unitary time evolution, although $|\cE_t\ra$\/ affects the evolution of organisms at time $t$\/ (see (\ref{Utaction})). 

For simplicity take all the $|\cE_t\ra$\/'s to have the same structure,
the superposition of two states 
$|\cE_t; B\ra$\/ and $|\cE_t; M\ra$\/:
\begin{equation}
 |\cE_t\ra = \mu_B |\cE_t; B\ra +\mu_M |\cE_t; M\ra, \label{envtsum}
 \end{equation}
 \xxx{envtsum}
 \yyy{Inferred from bxs 1,2 p HQ82 (p 49 is missing)}
where $\mu_B$\/ and $\mu_M$\/ are independent of $t$\/, and
 \bea
 \la \cE_t;B|\cE_t;B\ra  & = & \la \cE_t;M|\cE_t;M\ra  =1, \label{etBMnormalized} \\
 \la\cE_t; B|\cE_t; M\ra & = & 0, \label{BMtortho}\\
 |\mu_B|^2 + |\mu_M|^2 & = & 1, \label{Etnorm}  \\
 |\mu_B| >   |\mu_M|. \label{BgtM} 
 \eea 
 \xxx{etBMnormalized}
 \xxx{BMtortho}
 \xxx {Etnorm}
 \xxx{BgtM}
 \yyy{Bottom of p HQ146}

  The labels $B$\/ and $M$\/ stand, respectively, for ``Born-like'' and ``maverick'' states.  We will say more about this choice of terminology below; for now they are simply names for states satisfying (\ref{etBMnormalized})-(\ref{BgtM}).  
 
 The organisms ${\cO_B}$\/ and ${\cO_M}$\/ thrive only if they are in, respectively, Born-like and maverick environments, and decline otherwise.  Again being guided by the desire for simplicity,  we make the impact of the environment determinative and extreme.  If a species is alive at time $t$\/ in a favorable environment (${\cO_B}$\/ 
 in  $|\cE_t; B\ra$\/  or  ${\cO_M}$\/  in $|\cE_t; M\ra$\/), that species will be alive at time $t+1$\/. If a species is alive at time $t$\/ in an unfavorable environment (${\cO_B}$\/ 
 in  $|\cE_t; M\ra$\/  or  ${\cO_M}$\/  in $|\cE_t; B\ra$\/), that species will be extinct at time $t+1$\/.  
 
 For unitarity the dynamics must also include the possibility of transitions from states in which a species is extinct to states in which it is alive. To incorporate these transitions while allowing ``extinction'' to retain the meaning we want it to have, we introduce for each species a single alive state but $\Omega \gg 1$\/ extinct states. A species which becomes extinct progresses through a succession of extinct states until, after $\Omega$\/ time steps, it again becomes alive---$\Omega$\/ is essentially the Poincar\'{e} recurrence time.  This dynamics is compatible with unitarity, while restricting attention to states with $t < \Omega\/  $  allows us to model the desired (highly-simplified) evolutionary biology.  
 
 In detail: The state in which $B$\/ is alive is $|\cO_B; A\ra$\/, while the states in which $B$\/ is extinct are  $|\cO_B; X_j\ra$\/, $j=0,\ldots,\Omega-1$\/. Similarly,   
 $\cO_M$\/ is alive in $|\cO_M; A\ra$\/ and extinct in  $|\cO_M; X_j\ra$\/, $j=0,\ldots,\Omega-1$\/. Let $\wh{U}_t$\/ be the unitary operator that implements time evolution from time $t$\/ to time $t+1$\/.  We specify that $\wh{U}_t$\/ does not have any effect on the environment basis vectors  
 $|\cE_{t\pr}; B\ra$\/,  $|\cE_{t\pr}; M \ra$\/,   while the modification to the organism basis vectors effected by  $\wh{U}_t$\/  depends on the states of those vectors as well as on the environment at time $t$\/.   That is, at the next time step a species remains alive if the environment is definitely the type that is favorable for it, and becomes extinct if it is definitely in an unfavorable environment.  An extinct species in a favorable environment will remain extinct by transitioning to the the next extinct state, ``next''  being defined modulo $\Omega$\/.  An extinct species in an unfavorable environment will likewise remain extinct and transition to the next extinct state unless it is already in the final extinct state, in which case it will transition to the alive state. 
 
 The action of $\wh{U}_t$\/ on a basis of states of the $\cE_t$\/, $\cO_B$\/ and $\cO_M$\/ is thus given by 
  \bea
 \hspace*{-1.5cm}\wh{U}_t|\cE_t;B\ra |\cO_B;A\ra |\cO_M;A\ra&=&|\cE_t;B\ra|\cO_B;A\ra|\cO_M;X_0\ra   \label{Utaction}    \\   \nonumber 
   \hspace*{-1.5cm}\wh{U}_t|\cE_t;B\ra |\cO_B;A\ra |\cO_M;X_k\ra&=& |\cE_t;B\ra|\cO_B;A\ra \\ \nonumber   
                                                                           & &  \cdot \left(\rule[-.25cm]{0cm}{.5cm}|\cO_M;X_{k+1}\ra(1-\delta_{k,\Omega -1})+|O_M;A\ra\delta_{k,\Omega-1}\right) \\   \nonumber 
                                                                              \hspace*{-1.5cm}\wh{U}_t|\cE_t;B\ra |\cO_B;X_j\ra |\cO_M;A\ra &=& |\cE_t;B\ra \left(\rule[-.25cm]{0cm}{.5cm}   |\cO_B;X_{j+1}\ra (1-\delta_{j,\Omega-1}) +|\cO_B;X_0\ra\delta_{j,\Omega-1}\rule[-.25cm]{0cm}{.5cm}\right)  \\  \nonumber
                                                                            & & \cdot  |\cO_M,X_0\ra \\            \nonumber % \nonumber     
   \hspace*{-1.5cm}\wh{U}_t|\cE_t;B\ra |\cO_B;X_j\ra |\cO_M;X_k\ra&=& |\cE_t;B\ra %\cO_B;A\ra \\ \nonumber
                                                                            %& &  
                                                                            \left(\rule[-.25cm]{0cm}{.5cm}   |\cO_B;X_{j+1}\ra (1-\delta_{j,\Omega-1}) +|\cO_B;X_0\ra\delta_{j,\Omega-1}\rule[-.25cm]{0cm}{.5cm}\right)  \\  \nonumber
                                                                           & &  \cdot \left(\rule[-.25cm]{0cm}{.5cm}|\cO_M;X_{k+1}\ra(1-\delta_{k,\Omega -1})+|O_M;A\ra\delta_{k,\Omega-1}\right)   \\    \nonumber 
                                                                           \hspace*{-1.5cm}\wh{U}_t|\cE_t;M\ra |\cO_B;A\ra |\cO_M;A\ra &=&|\cE_t;M\ra|\cO_B;X_0\ra|\cO_M;A\ra   \\   \nonumber   \hspace*{-1.5cm}\wh{U}_t|\cE_t;M\ra |\cO_B;A\ra |\cO_M;X_k\ra&=& |\cE_t;M\ra|\cO_B;X_0\ra \\ \nonumber
                                                                           & &  \cdot \left(\rule[-.25cm]{0cm}{.5cm}|\cO_M;X_{k+1}\ra(1-\delta_{k,\Omega -1})+|O_M;X_0\ra\delta_{k,\Omega-1}\right)     \\   \nonumber 
                                                                              \hspace*{-1.5cm}\wh{U}_t|\cE_t;M\ra |\cO_B;X_j\ra |\cO_M;A\ra &=& |\cE_t;M\ra \left(\rule[-.25cm]{0cm}{.5cm}   |\cO_B;X_{j+1}\ra (1-\delta_{j,\Omega-1}) +|\cO_B;A\ra\delta_{j,\Omega-1}\rule[-.25cm]{0cm}{.5cm}\right)  \\  \nonumber
                                                                            & & \cdot  |\cO_M,A\ra    \\      \nonumber 
                                                                            \hspace*{-1.5cm}\wh{U}_t|\cE_t;M\ra |\cO_B;X_j\ra |\cO_M;X_k\ra&=& |\cE_t;M\ra %\cO_B;A\ra \\ \nonumber
                                                                            %& &  
                                                                            \left(\rule[-.25cm]{0cm}{.5cm}   |\cO_B;X_{j+1}\ra (1-\delta_{j,\Omega-1}) +|\cO_B;A\ra\delta_{j,\Omega-1}\rule[-.25cm]{0cm}{.5cm}\right)  \\  \nonumber
                                                                           & &  \cdot \left(\rule[-.25cm]{0cm}{.5cm}|\cO_M;X_{k+1}\ra(1-\delta_{k,\Omega -1})+|O_M;X_0\ra\delta_{k,\Omega-1}\right)                                                                                                                                                                                                                                                                                                                      
\eea
\xxx{Utaction}
\yyy{p HQ60}
  Indices $j$\/, $k$\/ run from $0$\/ through $\Omega-1$\/.
 
 Taking the inner products of the above states with a dual basis, and keeping in mind the orthogonality of  alive states to extinct states and of extinct states with different indices to each other, we obtain the matrix representation
 \be
  \wh{U}_t=  \hspace*{15cm} 
  \ee
$$
\left(\begin{array}{c |c | c | c | c | c | c | c}
    & \delta_{k,\Omega-1} & & & & & &  \\ \hline
  \delta_{m,0}& {\delta_{m,k+1}  \atop  (1-\delta_{k,\Omega-1}) } & & & & & & \\  \hline
 %                      &\cdot (1-\delta_{k,\Omega-1}) & & & & & & \\  \hline
                                              &  & &{ { [\delta_{l,j+1} \atop \cdot (1-\delta_{j,\Omega-1})}  \atop  { +  \delta_{l,0}\delta_{j,\Omega-1}] \atop  \cdot \delta_{k,\Omega-1} } } & & & &  \\   \hline
%                                              &  & &\cdot     & & & &  \\   
 %                                               &  & & +  \delta_{l,0}\delta_{j,\Omega-1}]  & & & & \\  
 %                                               &  & & \cdot \delta_{k,\Omega-1}  & & & & \\  \hline                                              
  &  & { {[\delta_{l,j+1} \atop \cdot (1-\delta_{j,\Omega-1})}  \atop {+ \delta_{l,0}\delta_{j,\Omega-1}] \atop \cdot \delta_{m,0} }}& 
  	{{[\delta_{l,j+1} \atop \cdot (1-\delta_{j,\Omega-1})} \atop {+\delta_{l,0} \delta_{j,\Omega-1}] \delta_{m,k+1} \atop (1-\delta_{k,\Omega-1})}}& & & & \\  \hline
 & & & & &   & \delta_{j,\Omega-1} &   \\ \hline
 & & & &  & & & { { [\delta_{m,k+1} \atop \cdot (1-\delta_{k,\Omega-1})}  \atop  { +  \delta_{m,0}\delta_{k,\Omega-1}] \atop  \cdot \delta_{j,\Omega-1} } }\\  \hline
 %                      &\cdot (1-\delta_{k,\Omega-1}) & & & & & & \\  \hline
                             & & & &     \delta_{l,0}              &  &{\delta_{l,j+1}  \atop  (1-\delta_{j,\Omega-1}) }  &   \\   \hline
%                                              &  & &\cdot     & & & &  \\   
 %                                               &  & & +  \delta_{l,0}\delta_{j,\Omega-1}]  & & & & \\  
 %                                               &  & & \cdot \delta_{k,\Omega-1}  & & & & \\  \hline                                              
& & & &  &   { {[\delta_{m,k+1} \atop \cdot (1-\delta_{k,\Omega-1})}  \atop {+ \delta_{m,0}\delta_{k,\Omega-1}] \atop \cdot \delta_{l,0} }}& &
  	{{[\delta_{m,k+1} \atop \cdot (1-\delta_{k,\Omega-1})} \atop {+\delta_{m,0} \delta_{k,\Omega-1}] \delta_{l,j+1} \atop (1-\delta_{j,\Omega-1})}} 
%  &  & & & & &  & \\   \hline
%  &  & & & & & & \\   \hline
%  &  & & & & & & \\   \hline
%  &  & & & & &  &
 %\caption{This is the table of $\wh{U}_t$\/}
 \end{array}\right) 
 $$ 
 \yyy{p HQ70}
 where $l$\/, $m$\/ are row indices and $j$\/, $k$\/ are column indices.  Blocks with no nonzero entries are left blank. 
 For example, the first entry in the first column is
 \be
 \la \cE_t;B|\la\cO_B;A|\la \cO_M;A|\wh{U}_t|\cE_t;B\ra |\cO_B;A\ra |
 \cO_M;A\ra
 =0, \label{Utmat11}
 \ee
 \xxx{Utmat11}
 \yyy{bx 1 p HQ-62}
 and the second entry in the first column is
 \be
 \la \cE_t;B|\la\cO_B;A|\la \cO_M;X_m|\wh{U}_t|\cE_t;B\ra |\cO_B;A\ra |
 \cO_M;A\ra
 =\delta_{m,0}. \label{Utmat21}
 \ee
 \xxx{Utmat21}
 \yyy{bx 2 p HQ-62} 
 From this representation it is straightforward to show that $\wh{U}_t$\/ is unitary.

 If we limit the time over which we track evolution to be  smaller than the Poincar\'{e} recurrence time, i.e., we take $T < \Omega-1$\/, then the indices $j$\/, $k$\/, 
 $l$\/ and $m$\/ will never be as large as $\Omega-1$\/ and  we can replace (\ref{Utaction})  with
 \bea
 \hspace*{-1.5cm}\wh{U}_t|\cE_t;B\ra |\cO_B;A\ra |\cO_M;A\ra&=&|\cE_t;B\ra|\cO_B;A\ra|\cO_M;X_0\ra    \label{prepoincareU}      \\   \nonumber     
     \hspace*{-1.5cm}\wh{U}_t|\cE_t;B\ra |\cO_B;A\ra |\cO_M;X_k\ra&=& |\cE_t;B\ra|\cO_B;A\ra  |\cO_M;X_{k+1}\ra \\   \nonumber 
     \hspace*{-1.5cm}\wh{U}_t|\cE_t;B\ra |\cO_B;X_j\ra |\cO_M;A\ra &=& |\cE_t;B\ra   |\cO_B;X_{j+1}\ra  |\cO_M,X_0\ra \\    \nonumber         % \nonumber     
   \hspace*{-1.5cm}\wh{U}_t|\cE_t;B\ra |\cO_B;X_j\ra |\cO_M;X_k\ra&=& |\cE_t;B\ra %\cO_B;A\ra \\ \nonumber
                                                                            %& &  
                                                                            |\cO_B;X_{j+1}\ra |\cO_M;X_{k+1}\ra  \\  \nonumber 
                                                                            \hspace*{-1.5cm}\wh{U}_t|\cE_t;M\ra |\cO_B;A\ra |\cO_M;A\ra &=&|\cE_t;M\ra|\cO_B;X_0\ra|\cO_M;A\ra   \\   \nonumber 
  \hspace*{-1.5cm}\wh{U}_t|\cE_t;M\ra |\cO_B;A\ra |\cO_M;X_k\ra&=& |\cE_t;M\ra|\cO_B;X_0\ra |\cO_M;X_{k+1}\ra    \\ \nonumber 
     \hspace*{-1.5cm}\wh{U}_t|\cE_t;M\ra |\cO_B;X_j\ra |\cO_M;A\ra &=& |\cE_t;M\ra   |\cO_B;X_{j+1}\ra  |\cO_M,A\ra       \\ \nonumber         
     \hspace*{-1.5cm}\wh{U}_t|\cE_t;M\ra |\cO_B;X_j\ra |\cO_M;X_k\ra&=& |\cE_t;M\ra |\cO_B;X_{j+1}\ra |\cO_M;X_{k+1}\ra.                                                                                                                                                                                                                                                                                                                     
 \eea
    \xxx{prepoincareU}  
    \yyy{Follows from p HQ60}
 
 Take the initial state at $t=0$\/, $|\Phi,0\ra$\/,  to be  one in which both species are alive:
 \be
 |\Phi,0\ra  =  |\cE \ra |\cO_B;A\ra |\cO_M;A\ra. \label{phit0} %\\
 %                & = & \prod_{t=0}^T \left(\rule[-.25cm]{0cm}{.5cm}|\cE_t;B\ra+|\cE_t;B\ra\right)|\cO_B;A\ra |\cO_M;A\ra 
 \ee
 \xxx {phit0}
 \yyy{bx 1 p HQ82}
 Evolution from one discrete time step to the next is implemented by
 \be
 |\Phi,t+1\ra=\wh{U}_t |\Phi,t\ra. \label{timestepevolution}
 \ee
 \xxx{timestepevolution}
 Using (\ref{envprod}), (\ref{envtsum}), (\ref{prepoincareU}), (\ref{phit0}) and (\ref{timestepevolution}), we find after the first  timestep
 \bea
 |\Phi,1\ra  & = & \left[\prod_{t=1}^T \left(\rule[-.25cm]{0cm}{.5cm}\mu_B |\cE_t;B\ra+\mu_M |\cE_t;M\ra \right)\right]  \nonumber  \\
& & 
 \left(\rule[-.25cm]{0cm}{.5cm} \mu_B | \cO_B; A \ra |\cO_M; X_0\ra |\cE_0; B \ra %\right.  \nonumber \\
 %& &  \left. \rule[-.25cm]{0cm}{.5cm}
 + \mu_M  | \cO_B;X_0 \ra | \cO_M;A \ra  | \cE_0;M \ra \right).   \label{Phit1}%\nonumber 
\eea
\xxx{Phit1}
\yyy{next-to-last bx p HQ83}
I.e., after one timestep there is a superposition of two states, in each of which only one species is alive.  This of course is a result of our having
chosen the two species to be incapable of surviving even a single generation in an unfavorable environment. After two timesteps,
\bea
 |\Phi,2\ra&=&\left[\prod_{t=2}^T \left(\rule[-.25cm]{0cm}{.5cm} \mu_B | \cE_t;B\ra+\mu_M |\cE_t;M\ra \right)\right] \label{Phit2} \\
                &  &\cdot \left(\rule[-.25cm]{0cm}{.5cm}(\mu_B)^2|\cO_B; A \ra | \cO_M; X_1 \ra |\cE_1; B\ra |\cE_0; B\ra+ 
                        \mu_B \mu_M| \cO_B;X_0 \ra | \cO_M;X_1 \ra | \cE_1;M \ra |\cE_0; B\ra\right. \nonumber \\
                &  &+ \left.\rule[-.25cm]{0cm}{.5cm}\mu_M \mu_B|\cO_B; X_1 \ra | \cO_M; X_0 \ra |\cE_1; B\ra |\cE_0; M\ra+ 
                       (\mu_M)^2 | \cO_B;X_1 \ra | \cO_M;A \ra | \cE_1;M \ra |\cE_0; M\ra\right) \nonumber                         
\eea
\xxx{Phit2}
\yyy{bxs 1,2 p HQ84}
 Proceeding by mathematical induction, we hypothesize that
 \bea   
  |\Phi,n\ra  & = & \left[\prod_{t=n}^T \left(\rule[-.25cm]{0cm}{.5cm}\mu_B |\cE_t;B\ra+\mu_M|\cE_t;M\ra \right)\right]      \label{Phitnindhyp}   \\  
                    &  &\cdot \left(\rule[-.25cm]{0cm}{.5cm}(\mu_B)^n|\cO_B; A \ra | \cO_M; X_{n-1} \ra |\cE_{n-1}; B\ra |\cE_{n-2}; B\ra \cdots |\cE_0; B\ra\right. \nonumber \\
                 & & + |\zeta_n\ra \nonumber \\
                &  &+ \left.\rule[-.25cm]{0cm}{.5cm}(\mu_M)^n|\cO_B; X_{n-1} \ra | \cO_M; A \ra |\cE_{n-1}; M\ra |\cE_{n-2}; M\ra \cdots |\cE_0; M\ra \right) \nonumber
 \eea
\xxx{Phitnindhyp}
\yyy{bx 1 p HQ88}
where $|\zeta_n\ra$\/ contains no factors of alive species, i.e., neither $|\cO_B;A\ra$\/  nor  $|\cO_M;A\ra$\/.  If (\ref{Phitnindhyp}) is true for some $n$\/, then, using 
(\ref{timestepevolution}), 
\bea
 |\Phi,n+1\ra  & = & \left[\prod_{t=n+1}^T \left(\rule[-.25cm]{0cm}{.5cm}\mu_B |\cE_t;B\ra+\mu_M |\cE_t;M\ra \right)\right] 
                               \wh{U}_n \left\{ \rule[-.5cm]{0cm}{1cm}\left(\rule[-.25cm]{0cm}{.5cm}|\mu_B \cE_n;B\ra+\mu_M |\cE_n;M\ra \right)\right. \label{indproof1}\\
                 &  &\cdot \left(\rule[-.25cm]{0cm}{.5cm}(\mu_B)^n |\cO_B; A \ra | \cO_M; X_{n-1} \ra |\cE_{n-1}; B\ra |\cE_{n-2}; B\ra \cdots |\cE_0; B\ra\right. \nonumber \\
                 & & + |\zeta_n\ra \nonumber \\
                &  &+ \left.\left.\rule[-.25cm]{0cm}{.5cm}(\mu_M)^n |\cO_B; X_{n-1} \ra | \cO_M; A \ra |\cE_{n-1}; M\ra |\cE_{n-2}; M\ra \cdots |\cE_0; M\ra \right)  \rule[-.5cm]{0cm}{1cm} \right\} \nonumber %
\eea
\xxx{indproof1}
\yyy{p.HQ-88, last eq.} 
% (\ref{Phit2}), (\ref{Phitnindhyp}) is true for
%$n=2\/$. 
and, using (\ref{prepoincareU}),
\bea
 |\Phi,n+1\ra  & = & \left[\prod_{t=n+1}^T \left(\rule[-.25cm]{0cm}{.5cm}\mu_B |\cE_t;B\ra+\mu_M |\cE_t;M\ra \right)\right] 
                                \label{indproof2}\\
                 &  &\left\{ \rule[-.5cm]{0cm}{1cm}\rule[-.25cm]{0cm}{.5cm}(\mu_B)^{n+1}|\cO_B; A \ra | \cO_M; X_{n} \ra |\cE_{n}; B\ra |\cE_{n-1}; B\ra |\cE_{n-2}; B\ra \cdots |\cE_0; B\ra 
\right. \nonumber \\
                &  &+ (\mu_B)^{n}\mu_M |\cO_B; X_0 \ra | \cO_M; X_{n} \ra |\cE_{n}; M\ra |\cE_{n-1}; B\ra |\cE_{n-2}; B\ra\cdots |\cE_0; B\ra \nonumber \\
                 & & +\wh{U}_n \left[ \rule[-.5cm]{0cm}{1cm}|\zeta_n\ra \left(\rule[-.25cm]{0cm}{.5cm}\mu_B| \cE_n;B\ra+\mu_M |\cE_n;M\ra \right)\right]\nonumber \\
                &  &+ (\mu_M)^n \mu_B |\cO_B; X_n \ra | \cO_M; X_0 \ra |\cE_{n}; B\ra |\cE_{n-1}; M\ra |\cE_{n-2}; M\ra \cdots |\cE_0; M\ra \nonumber \\
                &  &+ \left.\rule[-.25cm]{0cm}{.5cm}(\mu_M)^{n+1}|\cO_B; X_n \ra | \cO_M; A \ra |\cE_{n}; M \ra |\cE_{n-1}; M\ra |\cE_{n-2}; M\ra \cdots |\cE_0; M\ra   \rule[-.5cm]{0cm}{1cm} \right\} \nonumber %
\eea
\xxx{indproof2}
\yyy{p.HQ-89 (or maybe 88?), last eq.} 
By the induction hypothesis, $|\zeta_n\ra$\/ contains no factors of $|\cO_B;A\ra$\/ or $|\cO_B;M\ra$\/ and, as discussed above, $n < \Omega-1$\/, allowing us to use (\ref{prepoincareU}) rather than the full action (\ref{Utaction}) of $\wh{U}_t$\/. It follow that %$\wh{U}_n \left[ \rule[-.5cm]{0cm}{1cm}|\zeta_n\ra \left(\rule[-.25cm]{0cm}{.5cm}|\cE_n;B\ra+|\cE_n;M\ra \right)\right]$\/ 
$\wh{U}_n \left[ |\zeta_n\ra \left(|\cE_n;B\ra+|\cE_n;M\ra \right)\right]$\/  also contains no factors of  $|\cO_B;A\ra$\/ or $|\cO_B;M\ra$\/, nor does $|\zeta_{n+1}\ra$\/ defined as
\bea
                |\zeta_{n+1}\ra=&  &(\mu_B)^n \mu_M  |\cO_B; X_0 \ra | \cO_M; X_{n} \ra |\cE_{n}; M\ra |\cE_{n-1}; B\ra |\cE_{n-2}; B\ra\cdots |\cE_0; B\ra \label{zetanplus1def} \\
                 & & +\wh{U}_n \left[ \rule[-.5cm]{0cm}{1cm}|\zeta_n\ra \left(\rule[-.25cm]{0cm}{.5cm}\mu_B |\cE_n;B\ra+\mu_M |\cE_n;M\ra \right)\right]\nonumber \\
                &  &+(\mu_M)^n \mu_B |\cO_B; X_n \ra | \cO_M; X_0 \ra |\cE_{n}; B\ra |\cE_{n-1}; M\ra |\cE_{n-2}; M\ra \cdots |\cE_0; M\ra. \nonumber 
\eea
\xxx{zetanplus1def}
\yyy{p. HQ-90, box1}
%is also a vector with no factors of $|\cO_B;A\ra$\/ or $|\cO_B;M\ra$\/.  
So,  using 
(\ref{zetanplus1def}) to write  (\ref{indproof2}) as 
\bea
 |\Phi,n+1\ra  & = & \left[\prod_{t=n+1}^T \left(\rule[-.25cm]{0cm}{.5cm}\mu_B |\cE_t;B\ra+\mu_M|\cE_t;M\ra \right)\right] 
                                \label{indproofdone}\\
                 &  &\left(\rule[-.25cm]{0cm}{.5cm}(\mu_B)^{n+1} |\cO_B; A \ra | \cO_M; X_{n} \ra |\cE_{n}; B\ra |\cE_{n-1}; B\ra |\cE_{n-2}; B\ra \cdots |\cE_0; B\ra 
\right. \nonumber \\
                 & & + |\zeta_{n+1}\ra \nonumber \\
                &  &+ \left.\rule[-.25cm]{0cm}{.5cm}(\mu_M)^{n+1}  |\cO_B; X_{n} \ra | \cO_M; A \ra |\cE_{n}; M\ra |\cE_{n-1}; M\ra |\cE_{n-2}; M\ra\cdots |\cE_0; M\ra \rule[-.25cm]{0cm}{.5cm}  \right)\nonumber %
\eea
\xxx{indproofdone}
\yyy{p.HQ-90, box 2.}
and comparing with (\ref{Phitnindhyp}), we see that the induction hypothesis is true for $n+1$\/ if it is true for $n$\/  (since $|\zeta_{n+1} \ra$\/ satisfies the same conditions as required for $|\zeta_n\ra$\/) and so,  with (\ref{Phit1}) and  (\ref{Phit2}), is  true for all $n$\/.

For $n>0$\/ there is one branch of the Schr\"{o}dinger-picture state (\ref{Phitnindhyp}) in which the species $\cO_B$\/ is alive and $\cO_M$\/ is extinct, and another distinct branch in which $\cO_M$\/ is alive and $\cO_B$\/ is extinct. In all other branches neither species is alive. The weights for 
$\cO_B$\/ and $\cO_M$\/ being alive are the respective square norms of these branches:
\bea
W_B & = & |\mu_B|^{2n} ,\label{WBlabel}\\
W_M & = & |\mu_M|^{2n}.
\eea
\yyy{bx 2 p HQ91, last bx p HQ90}
Both $|\mu_B|$\/ and  $|\mu_M|$\/  are less than unity, and we've specified that $|\mu_B| > |\mu_M|$\/.  So as the number of generations $n$\/ of biological evolution increases, both $W_B$\/ and $W_M$\/ will decrease, with $W_M$\/ dropping below the preclusion threshold first.  That is, for at least some period of time the species that is fit to survive in a higher-weight (``more probable'') environment will be present while the species fit to survive in a lower-weight (``less probable'') environment will be extinct.
 
\subsection{Born-like and maverick states} \label{subsecBornlikemaverickstates}

\xxx{subsecBornlikemaverickstates}

If we now base  a model of the environment on the same collection of $N$\/ spins described in Sec. \ref{SecIncompatibility}, we can show that the sum of branches with relative frequencies for spin-up and spin-down near the Born-rule values will constitute a state with norm increasing towards unity as $N$\/ becomes large, while the remaining ``maverick'' branches, i.e., those with relative frequencies far from the Born-rule values, will constitute a state with norm approaching zero for large $N$\/.  We can therefore identify the respective sums at time $t$\/ of Born-like branches and maverick branches with $\mu_B|\cE_t;B\ra$\/ and $\mu_M |\cE_t;M\ra$\/ of Sec. \ref{biolevolhardprecl}, since these will satisfy the conditions $\la\cE_t;B|\cE_t;M\ra=0$\/ and %$\la\cE;B|\cE;B\ra > \la\cE;M|\cE;M\ra$\/. 
$|\mu_B| > |\mu_M|$\/. Applying the results of Sec. \ref{biolevolhardprecl} we conclude that species that thrive in a Born-like environment (for purposes of visualization,  consider an organism that eats spins and must consume up-spins and down-spins in a ratio close to $p:q$\/) will persist in the presence of a preclusion threshold that would lead organisms requiring a far-from-Born-rule environment to become extinct.

Referring to %(\ref{SPEvmeas}), 
(\ref{productstate})-(\ref{rsum}), define a state composed of branches\footnote{We will be identifying $|\wti{\cE}_t;B\ra$\/, $|\wti{\cE}_t;M\ra$\/, up to norm, 
with $|\cE_t;B\ra$\/, $|\cE_t;M\ra$\/; see (\ref{identifications}). As discussed in Sec.~\ref{biolevolhardprecl}, $|\cE_t\ra$\/ represents a physical object  distinct  from %$|\cE_t\pr\ra$\/, 
$|\cE_{t^{\prime}}\ra$\/, $t \neq t\pr$\/,  so strictly speaking we should write in the rest of this section $ |B_{i_1,\ldots,i_N}\ra_t$\/, where 
$ |B_{i_1,\ldots,i_N}\ra_t$\/ is defined by (\ref{SPEvmeas}) with $|\cS\pup;\al_{i^{(p)}}\ra$\/ replaced by $|\cS\pup;\al_{i^{(p)}}\ra_t$\/.  } with relative frequencies near the Born-rule values,
\bea \label{EBdef}
|\wti{\cE}_t;B\ra &=&  \sum_{i_1=1}^2 \cdots \sum_{i_N=1}^2 |B_{i_1,\ldots,i_N}\ra \nonumber \\
                 & &   \theta\left(r_1(i_i,\ldots,i_N)-N(p-\delta)\right) %\nonumber \\
                  \theta\left(N(p+\delta)-r_1(i_i,\ldots,i_N)\right), 
\eea
\xxx{EBdef}
\yyy{bx. 2, p. HQ 11}
and a state composed of the remaining branches with relative frequencies farther from the Born-rule values,
\bea  \label{EMdef}
|\wti{\cE}_t;M\ra & = & |\Psi\ra -|\wti{\cE}_t;B\ra \nonumber \\
&=&  \sum_{i_1=1}^2  \cdots \sum_{i_N=1}^2 |B_{i_1,\ldots,i_N}\ra \nonumber \\
                 & &   \left[\rule[-.25cm]{0cm}{.5cm}\widetilde{\theta}\left(N(p-\delta)-r_1(i_i,\ldots,i_N)\right) %\nonumber \\
                  + \widetilde{\theta}\left(r_1(i_i,\ldots,i_N)-N(p+\delta)\right) \right],
\eea
\xxx{EMdef}
\yyy{ bx2 pHQ-11, and using bx 1 p HQ 27}
where 
\be
0 < \delta < \min(p, q),\label{deltarange}
\ee
\xxx{deltarange}
and
\bea \label{thetadef}
\theta(x)&=&0, \hspace*{1cm}x < 0 \nonumber\\
              & = & 1, \hspace*{1cm}x = 0 \nonumber\\
              & = & 1, \hspace*{1cm}x >  0, 
\eea
\xxx{thetadef}
\yyy{last bx p HQ21}
\bea \label{thetatildefef}
\wti{\theta}(x)&=&0, \hspace*{1cm}x < 0, \nonumber\\
              & = & 0, \hspace*{1cm}x = 0, \nonumber\\
              & = & 1, \hspace*{1cm}x > 0, \nonumber
\eea
\xxx{thetatildedef}
\yyy{first bx p HQ22}
so
\be
\left(\theta(x)\right)^2=\theta(x),\label{thetasquared}
\ee
\xxx{thetasquared}
\be
\left(\widetilde{\theta}(x)\right)^2=\widetilde{\theta}(x),\label{thetatildesquared}
\ee
\xxx{thetatildesquared}
\be
\theta(x)+\wti{\theta}(-x)=1,\label{thetaplusthetatilde}
\ee
\xxx{thetaplusthetatilde}
\yyy{bx 3 p HQ-22}
\be
\theta(x)\wti{\theta}(-x)=0.\label{thetatimesthetatilde}
\ee
\xxx{thetatimesthetatilde}
\yyy{bx 4 p. HQ-22}
From (\ref{SPEvmeas}),
\be
\la B_{j^{(1)},\ldots,j^{(N)}}|B_{i^{(1)},\ldots,i^{(N)}}\ra=|c_1|^{2 r_1(i^{(1)},\ldots,i^{(N)})} |c_2|^{2 r_2(i^{(1)},\ldots,i^{(N)})}
\prod_{p=1}^N \delta_{j^{(p)},i^{(p)}}    \label{BMdotprod}
\ee 
\xxx{BMdotprod}
\yyy{bx 5 p HQ21}
Using (\ref{EBdef}), (\ref{EMdef}),  (\ref{thetatimesthetatilde}) and (\ref{BMdotprod}), we verify that
\be
\la \wti{\cE}_t;B|\wti{\cE}_t;M\ra=0. \label{EBEMorthogonal}
\ee
\xxx {EBEMorthogonal}
\yyy{bx 1 p HQ-27}
From (\ref{ridef}) we see that $r_i(i^{(1)},\ldots,i^{(N)})$\/ takes integer values in the range $[0,N]$\/,
\be
0 \leq  r_i(i^{(1)},\ldots,i^{(N)})  \leq N. \label{rangeofri}
\ee
\xxx{rangeofri}
\yyy{bx1 p JQ-24}
Therefore
\be
\sum_{n=0}^N \delta_{n,r_i(i^{(1)},\ldots,i^{(N)})}    = 1.  \label{sumnri} %,r_i(i^{(1)},\ldots,i^{(N)}) = 1. %\label{sumnri}
\ee
\xxx{sumnri}  
\yyy{bx 2 p HQ-23}
Using (\ref{EBdef}), (\ref{thetasquared}) and (\ref{BMdotprod})
\bea 
\la \wti{\cE}_t;B|\wti{\cE}_t;B\ra &= & \sum_{i^{(1)}=1}^2 \cdots \sum_{i^{(N)}=1}^2   \label{EBsq1} \\
                             &    &  \theta\left(r_1(i_i,\ldots,i_N)-N(p-\delta)\right) \theta\left(N(p+\delta)-r_1(i_i,\ldots,i_N)\right)  \nonumber \\
                  	    &    & |c_1|^{2r_1(i^{(1)},\ldots,i^{(N)})}    |c_2|^{2r_2(i^{(1)},\ldots,i^{(N)})}   \nonumber \\
	    		    & = & \sum_{i^{(1)}=1}^2 \cdots \sum_{i^{(N)}=1}^2 \sum_{n=0}^N \delta_{n,r_i(i^{(1)},\ldots,i^{(N)})}  \label{EBsq2}\\
                             &    &  \theta\left(r_1(i_i,\ldots,i_N)-N(p-\delta)\right) \theta\left(N(p+\delta)-r_1(i_i,\ldots,i_N)\right)  \nonumber \\
                  	    &    & |c_1|^{2r_1(i^{(1)},\ldots,i^{(N)})}    |c_2|^{2r_2(i^{(1)},\ldots,i^{(N)})}   \nonumber 			    
\eea
\xxx{EBsq1,EBsq2}
\yyy{bx2, 4 p HQ-28}
using (\ref{sumnri})
\bea
                             & = & \sum_{i^{(1)}=1}^2 \cdots \sum_{i^{(N)}=1}^2 \sum_{n=0}^N \delta_{n,r_i(i^{(1)},\ldots,i^{(N)})} \label{EBsq3}\\
                             &    &  \theta\left(n-N(p-\delta)\right) \theta\left(N(p+\delta)-n\right)  \nonumber \\
                  	    &    & |c_1|^{2n}    |c_2|^{2(N-n)}   \nonumber \\
	    		     & = & \sum_{n | N(p-\delta) \le n \le N(p+\delta)}   {N \choose n}      |c_1|^{2n}    |c_2|^{2(N-n)}.             	
\eea
\xxx{EBsq3}
\yyy{bx4, p HQ-28}
Using (\ref{pqdef}), rewrite this as
\be
\la \wti{\cE}_t;B|\wti{\cE}_t;B\ra = \sum_{n | -\delta\sqrt{\frac{N}{pq}} \le \frac{n-Np}{\sqrt{Npq}} \le \delta\sqrt{\frac{N}{pq}} }   {N \choose n}      p^{n}    q^{N-n}     \label{sumforEBsq}
\ee
\xxx{sumforEBsq} 
\yyy{bx 1 p HQ-151}

Following the approach of Sec. {\ref{SecModel}, we use (\ref{DprAepslim}), (\ref{lolimerf}) and (\ref{epsdef})  in (\ref{sumforEBsq}) to obtain
$$
\la \wti{\cE}_t;B |\wti{\cE}_t;B  \ra > 1-\eps,\hspace*{1cm} 
       N > \max\left(N_D(\eps_D,A), \frac{A^2 pq}{\delta^2}\right), 
       \hspace*{2mm} A>A_e(\eps_e), \hspace*{2mm}
       \eps > \eps_D+\eps_e,  \hspace*{2mm} 
 $$
\be
\eps_D, \eps_e, \eps >0 \label{sqnormEBlim}
\ee
\xxx{sqnormEBlim}
\yyy{bx 1 p. WC-73}
From (\ref{EBdef}), (\ref{EMdef}) and (\ref{EBEMorthogonal}) 
\be
\la \wti{\cE}_t;B| \wti{\cE}_t;B\ra+\la\wti{\cE}_t;M|\wti{\cE}_t;M\ra = \la \Psi | \Psi \ra = 1. \label{normsums}%\nonumber\\                                                      
\ee
\xxx{normsums}
\yyy{bx 1 p HQ-153} 
Using (\ref{sqnormEBlim}) and (\ref{normsums}),
$$
\la \wti{\cE}_t;M |\wti{\cE}_t;M  \ra < \eps,\hspace*{1cm} 
       N > \max\left(N_D(\eps_D,A), \frac{A^2 pq}{\delta^2}\right), 
       \hspace*{2mm} A>A_e(\eps_e), \hspace*{2mm}
       \eps > \eps_D+\eps_e,  \hspace*{2mm} 
 $$
\be
\eps_D, \eps_e, \eps >0 \label{sqnormEMlim}
\ee
\xxx{sqnormEMlim}
\yyy{bx 3 p. WC-73} 
 So, if we make the identifications, for all $t$\/,
\bea
|\wti{\cE}_t;B\ra&=& \mu_B|\cE_t;B\ra \nonumber\\
|\wti{\cE}_t;M\ra&=& \mu_M|\cE_t;M\ra \label{identifications}
\eea
\xxx{identifications}
\yyy{see p WC-74}
we see from (\ref{etBMnormalized}), (\ref{sqnormEBlim}) and (\ref{sqnormEMlim}) 
%(\ref{normsums}), (\ref{EBsqNlim}) and (\ref{EMsqNlim}) 
that (\ref{BMtortho}), (\ref{Etnorm}) and (\ref{BgtM}) are satisfied for any $\delta$\/ satisfying (\ref{deltarange}) provided $N$\/ is large enough so that $\eps < 1/2$\/.  

The results of  Sec. \ref{biolevolhardprecl} therefore apply, and we can conclude that species fit to survive in environments $|\cE_t;B\ra$\/ 
with relative frequencies near the Born-rule value---and one aspect of fitness to survive in an environment can be possessing a sense of subjective probability consistent with that environment---will have an advantage over species fit to survive in maverick environments $|\cE_t;M\ra$\/.\footnote{The reader may object that  in focusing on the states defined in (\ref{EBdef}) and (\ref{EMdef}) we are performing the same sort of unjustified grouping that we objected to in Sec. \ref{SecIncompatibility}.  It seems clear that this objection would be dispelled by an analysis in the Heisenberg picture, where the focus would be on the operators representing $\cO_B$\/ and $\cO_M$\/.  The ``preferences''  of $\cO_B$\/ and $\cO_M$\/ for different ranges of ratios of up-spins and down-spins, i.e., the different transformations induced by environments with different  ratios,  would group the branches in (\ref{EBdef})  and (\ref{EMdef})  just as in Sec. \ref{SecModel} the different transformations of $\cF$\/ due to  different ratios of observers' perceptions leads to a grouping, with the parameter $\delta$ playing a role analogous to that played by  \/$1/\nu$\/ .}

\subsection{Experimental evidence for the biological evolution of subjective judgements of probability of ordinary probabilistic events}\label{SecExperiments}
\xxx{SecExperiments}

Is there evidence that subjective probability is in fact a product of evolution and is to some extent innate to individual organisms?  A substantial body of research would seem to support this idea.

Human infants less than a year in age show surprise at the occurrence of improbable events, for example the outcomes of lottery-like games.
``\ldots 12-month-olds have rational expectations about
the future based on estimations of event possibilities, without the
need of sampling past experiences. \ldots Our results suggest that at the onset of human decision
processes the mind contains an intuition of elementary probability
that cannot be reduced to the encountered frequency of events or
elementary heuristics. \ldots
We presented
movies in which three identical objects and one different in color and 
shape bounced randomly inside a container with an open
pipe at its base, as in a lottery game \ldots [Subsequently], 
an occluder hid the container and
one object, either one of the three identical objects (probable
outcome) or else the different one (improbable outcome), exited
from the pipe.
\ldots
Infants had no
information about frequency distributions of actual outcomes, so
their reactions could not be primed by previous experience.
\ldots
Despite the complexity of the task and the lack of habituation,
infants looked significantly longer when they witnessed the
improbable outcome.
\ldots
This result suggests that infants do not need to experience
outcome frequency to respond to probabilities.
\ldots 
%Together, t
These experiments show that just as infants expect
that future events will respect physical constraints \ldots%(16) 
they
also expect that in the future the most likely outcome will
occur. Because no frequency information about actual outcomes
was provided, these expectations are grounded on
intuitions about single event probabilities based on future
possibilities. Undoubtedly, infants respond to frequencies \ldots. However, our experiments show that the origin of the
concept of probability cannot be reduced to experiencing
frequencies \cite{Teglasetal07}.''%{\bf [cite T\'{e}gl\'{a}s et al. (2007)]}

Results in a similar experiment  
``suggest that not only were infants {\em surprised}\/ 
at improbable outcomes, but also that they {\em anticipated}\/  the
occurrence of a probable future event, even without any previous
experience with them.
\ldots Infants
can reason about single-case probabilities without previous experience
with the outcomes of the scenes as early as 12 months of
age. This result adds compelling further evidence to the finding
that single-case probabilities are meaningful for young humans
when they can represent and track possible outcomes \ldots  A
well-known position in evolutionary psychology holds that
humans can understand probabilities only as collections of experienced
events \ldots According to this perspective, our intuitions about the
future are entirely dependent on our experience of the past.
Instead, we showed that infants have no difficulty in reacting to
single-case probabilities despite having no information about the
past frequencies of the outcomes \cite{TeglasBonatti16}.''%{\bf [cite T\'{e}gl\'{a}s and Bonatti (2016)]}

Infants also show surprise at mismatch between the statistical properties of populations of items with the statistical properties of samples randomly drawn from those populations. They can use the observed statistics of populations of items to guide actions, and are sensitive to the presence or absence of random sampling. 
``Human learners make inductive inferences based on small amounts
of data: we generalize from samples to populations and vice versa.
The academic discipline of statistics formalizes these intuitive
statistical inferences. What is the origin of this ability? We report
six experiments investigating whether 8-month-old infants are
`intuitive statisticians.' Our results showed that, given a sample,
the infants were able to make inferences about the population
from which the sample had been drawn. Conversely, given information
about the entire population of relatively small size, the
infants were able to make predictions about the sample. 

`` \ldots One important assumption in a statistical inference task,
using probability or heuristics, is the assumption of random
sampling. It is only under conditions of random sampling that the
inference is warranted. If the learner has evidence that she is not
receiving a random sample from the population, she can no
longer use the statistical information in the sample to make
guesses about the overall population.

``Recent studies in our laboratory suggest that 11-month-old
infants are sensitive to sampling conditions.  \ldots Infants were
given evidence that the random sampling assumption had been
violated. The experimenter first expressed a preference for one
type of ping-pong balls, say red ones, by showing the infant a
small container with both types of ping-pong balls (red and
white) and selectively picking up only the red ones and placing
them in another container. On the test trials, the experimenter
looked into the box while pulling out the ping-pong balls, i.e., she
had visual access to the content of the box. If infants were
sensitive to the fact that the random sampling assumption had
been violated, their looking times should no longer be predicted
by which sample was more probable given the content of the box.
Instead their looking times should be predicted by whether the
sample was consistent with the experimenter's preference or not.
That was exactly what we found; infants looked longer when the
experimenter pulled out a sample that was inconsistent with her
expressed preference, regardless of the content of the box. In the
third condition, the experimenter expressed a preference but she
was blindfolded during the sampling process. The infants were
able to integrate these two sources of information, and their
looking times were once again predicted by the content of the
box (F. Xu and S. Denison, unpublished data). These results
suggest that infants engaged in a rather sophisticated form of
statistical inference in this task, and their looking time patterns
on the test trials were not simply a matter of matching the sample
to the overall population in terms of distribution.

``The present studies provide evidence that early in development
infants are able to use a powerful statistical inference
mechanism for inductive learning. They can make generalizations
about a population based on a sample, and conversely, they
can make predictions about a sample given information about a
population. This ability for performing intuitive statistics develops
early and in the absence of schooling or explicit teaching. It
may be the roots of later acquisition of statistical principles, in
both the course of developing an understanding of scientific
inquiry and learning about probabilistic reasoning and statistics\cite{XuGarcia08}.''%{\bf [Cite Xu and Garcia (2008)]}

Similar results have been obtained in studies of nonhuman primates; see, e.g.,\cite{HanusCall14,Rakoczyetal14, Tecwynetal17,DePetrilloRosati19}.  For a recent review of research in this area, see \cite{DenisonXu19}.%{\bf [Cite Denison and Xu (2019)].}

\section{Summary}\label{SecSumm}
\xxx{SecSumm}

In brief: Hard preclusion, applied to the Heisenberg-picture ontology of Everett quantum mechanics, explains the nonobservation of extraordinary probabilistic phenomena, e.g., macroscopic entropy decrease. The evolution of organisms fit to maverick environments, i.e., environments with statistics far from Born-rule values, is an extraordinary probabilistic phenomenon and is precluded; nonprecluded species evolve subjective judgements of probability consistent with Born-rule values. 

In more detail:

\begin{enumerate}
\item We show that hard preclusion cannot account for the nonobservation of extraordinary probabilistic phenomena  (Sec.~\ref{SecIntro}) if the ontology of the Everett interpretation implied by the Schr\"{o}dinger picture is employed (Sec.~\ref{SecIncompatibility}).
\item We show that, if the Everett-interpretation ontology implied by the Heisenberg picture (Sec.~\ref{SecEQMHP}) is employed, hard preclusion can account in a straightforward and objective manner for the nonobservation of extraordinary probabilistic phenomena. % such as macroscopic violations of the second law of thermodynamics (Sec.~\ref{SecIntro}). 
As a model of extraordinary probabilistic phenomena we analyze an ensemble of spins, demonstrating for sufficiently large ensemble size the preclusion of relative-frequency measurements differing from those given by the Born rule and, as a by-product, extending the results of \cite{Rubin03} for the case of infinite ensemble size (Sec.~\ref{SecModel}).  A model showing the consistency of preclusion when applied to observations of contingent events is also presented (Sec.~\ref{SecPreclusionConsCon}).
\item We argue that the long time scales involved in biological evolution render the quantum weights for the survival of species with subjective judgements of probability differing greatly from those given by the Born rule small enough to be subject to preclusion, thus accounting for the perception, by surviving species, of ordinary probabilistic phenomena in a manner consistent with the  Principal Principle 
~(Sec. \ref{SecConnection}).
\item A proof-of-concept model of natural selection relying on hard preclusion rather than objective probability is presented 
(Sec.~\ref{biolevolhardprecl}).  A specialization of this model to two species existing in a %superposition of 
spin-ensemble environment shows that species that thrive in Born-like states can survive preclusion thresholds that drive species that thrive in maverick states to extinction
 (Sec~\ref{subsecBornlikemaverickstates}).    
\item We review experimental evidence that subjective probability is innate, a  product of biological evolution (Sec.~\ref{SecExperiments}).
\end{enumerate}

\section{Discussion}\label{SecDisc}
\xxx{SecDisc}

Theories of probability have been classified as either ``objective'' or ``subjective''  \cite{deFinetti17}. The present theory is an objective theory in that it is based on the objective existence of both quantum weights and the preclusion threshold, both features of the physical world.   In particular it is a matter of objective fact whether or not the weight for a quantum event is so small that the event is precluded.  The probabilities that are assigned to ordinary probabilistic phenomena, however, reside purely in the minds of organisms performing probabilistic judgments, e.g., the infants in the experiments described in Sec. \ref{SecExperiments}. As already pointed out, such judgements are themselves objective phenomena. %In the words of de Finetti, 
``The subjective opinion, as something known by the individual under consideration is, at least in this sense, something objective and can be a reasonable subject  of a rigorous study \cite[p. 5]{deFinetti17}.''  These subjective opinions, in turn, result from the objective process of biological evolution based on quantum mechanics with preclusion.  

The idea that subjective probability is at least to some degree innate and a product of evolution has been entertained by at least one subjectivist. Koopmans writes that ``probability derives directly from the intuition \ldots probability as well as logic may be derived by race experience through the process of evolution\cite{Koopmans40}.''   But, in the usual description, biological evolution
 is a process guided by objective probability. Without objective probability---or some other objective feature of the world, such as the hard preclusion employed here---no explanation is provided as to why one species survives and another becomes extinct, or why one trait is preserved and another extinguished.  This point is discussed further below.

%In decision-theoretic theories of probability in the Everett interpretation\cite{Deutsch99,DeWitt98, Polley00,Wallace02,Wallace03,Wallace07,Forrester07,Wallace10,SebensCarroll18}, the existence of decision-making agents satisfying certain nonprobablistic conditions of rationality is assumed.  The laws of quantum mechanics are then argued to constrain the agents to act as if they were maximizing probabilistic expectation values. But rational agents come about through biological evolution, and without objective probability or some  surrogate such as preclusion the fact that the decision-making agents exist in the first place is unexplained. 

The hard-preclusion theory as presented above has a ``two-level ontology:''\footnote{The theory thus calls to mind Heisenberg's  ``potentia'' \cite{Heisenberg58, Heisenberg74}.  The place of this notion in quantum mechanics has been explored extensively (see, e.g., \cite{Fleming92,Pangle14,Bschir16,BishopBrenner17,Jaeger17,Jaeger17b,Kastner18,Kistler18,Sanders18}) but, to the best of my knowledge, always in terms of a single possibility being actualized stochastically. In the present theory, all possibilities with weights above the preclusion threshold are actualized deterministically.}  an underlying level of Heisenberg-picture operators which evolve  continuously; and a level of events which either occur or fail to occur as the weights associated with them are respectively above or below the preclusion threshold \cite{Rubin10}.  In this it is like any other quantum theory in which Heisenberg-picture operators correspond to events \cite{DeutschHayden00,Rubin01, Rubin02,Rubin03,Rubin04, Rubin06, HewittHorsmanVedral07a,HewittHorsmanVedral07b,Forrester07,Rubin11,Deutsch12,Bedard20,KuypersDeutsch20}, except that in the  usual theories the preclusion threshold is (implicitly)
set to a value of zero.\footnote{E.g., neither of the terms on the right-hand side of (\ref{bt}) vanishes regardless of the initial state,  but if either has zero matrix element in the initial state the corresponding outcome will not occur.}   Combining the two levels into one may be possible by introducing changes to the Heisenberg-picture equations of motion for the operators, e.g.  a Heisenberg-picture version of the Schr\"{o}dinger-picture  ``snap-to'' rule proposed in  BHZ. %\footnote{BHZ motivate preclusion from hypothesized discreteness of Hilbert space, which in turn is motivated by the granularity of spacetime suggested by quantum-gravitational considerations\cite{Hsu20}.}
 Whether or not this is done, however, hard preclusion 
is not simply an interpretation of quantum mechanics but a modification to it.

   Now, one goal of the Everett program has been  
   ``to take the mathematical formalism of quantum mechanics as it stands 
   without adding anything to it\cite{DeWitt70}.''
If we find it necessary to modify the formalism,\footnote{Van Esch\cite{VanEsch07}, Barrett\cite{Barrett17} and Vaidman\cite{Vaidman20} argue that  adding assumptions of some sort to the formalism is  a requirement  if one is  to obtain probabilities.}  is  the effort worth it?  The answer to that question remains as simple and compelling as ever:  The Everett interpretation, by virtue of eliminating the insistence on a single outcome to quantum measurement,   
avoids the nonlocality \cite{Everett57,Page82,Stapp85,DeWitt98,Tipler00,DeutschHayden00,Rubin01,Bacciagaluppi02,TimpsonBrown02,Vaidman02,HewittHorsman09,Rubin11,Deutsch12,Vaidman15}  that Bell's theorem assures us must be present in any single-outcome quantum theory \cite{Bell64,Bell75,Bell81,Bell90,Norsen06,Blaylock10,Maudlin14,Myrvoldetal20}.\footnote{Weissman\cite{Weissman99} introduces nonlinear modifications to the Schr\"{o}dinger equation but retains multiple outcomes, thereby retaining locality.}

%{\bf \textcolor{green}{\\ \noindent START NEW MOTIVATION\\}}

The Everett interpretation thereby being an eminently worthwhile enterprise, it is all the more important to make as clear as possible the reasons for incorporating preclusion into Everett quantum mechanics. 
%What motivations are there for believing that preclusion is in fact a law of nature?
There are two  motivations for believing that preclusion is in fact a real physical process.

The first of these motivations is the one which has been guiding our presentation throughout this paper: {\em Everett quantum mechanics with preclusion provides an explicitly-specified, noncircular explanation for probabilistic phenomena}\/.  To recapitulate, events with quantum-mechanical weight below the preclusion threshold do not occur.  In particular, events corresponding to macroscopic entropy decrease do not occur. But in addition the biological evolution of organisms fit to non-Born-rule ``maverick'' environments does not occur, explaining why those organisms 
that do evolve
%the evolution of which does occur 
possess traits, including the trait of subjective probability, fit to Born-rule environments.  The full gamut of probabilistic phenomena, nonobservation of extraordinary probabilistic phenomena as well as subjective judgements of ordinary probabilistic phenomena, is explained.

Deutsch has stated: ``A scientific {\em explanation}\/ is a statement of what is there in reality, and how it behaves and how that accounts for the explicanda\cite{Deutsch16}.''  We agree with this idea of %point of view regarding 
the meaning of explanation, from which it follows that a necessary condition for something to be the explanation of a physical phenomenon is that the thing that provides the explanation is something that exists objectively.   From this viewpoint, let us briefly examine some prominent approaches to Everettian probability.%\footnote{A detailed review of the full range of approaches to Everettian probability is outside the scope of the present paper.}

\begin{itemize}

\item 
In the Deutsch-Wallace decision-theoretic approach\cite{Deutsch99,DeWitt98, Polley00,Wallace02,Wallace03,Wallace07,
%Forrester07,
Wallace10
%,SebensCarroll18
}, the existence of decision-making agents satisfying certain nonprobablistic conditions of rationality is assumed.  The laws of quantum mechanics are then argued to constrain the subjective judgements of the agents to act as if they were maximizing probabilistic expectation values. But rational agents come about through biological evolution, and without objective probability or some  surrogate such as preclusion the fact that the decision-making agents exist in the first place is 
unexplained.\footnote{Wallace argues that objective as well as subjective probabilities are present in his version of the decision-theoretic approach\cite[p. 249] {Wallace12}, but this claim is questioned by Brown and Ben Porath: ``If agent-independent
chances existed, then there would be more to Everettian ontology
than just the universal wavefunction, contrary to the aims of the approach\ldots  Note that the probabilities in the argument are 
 %the exercise is at
%heart the application of a sophisticated variant of the principle of indifference
%based on symmetry. This makes the nature of quantum probabilities essentially
%something Laplace would recognise
 credences\ldots\cite{BrownBenPorath20}.''   Even if the approach somehow gave rise to something we would consider to be objective probability, it would only do so in a circular fashion. }
  If we take as given the existence of the organisms and that they follow the specified rules of rationality\footnote{This in fact seem to be the approach
  Deutsch is taking. ``[No] analysis from within physics [could] ever decide what is rational, or what is within the scope of reason\cite{Deutsch99}.''  We would argue that rational behavior, being a type of behavior leading to the  survival of the species, could come about through biological evolution guided by preclusion, along the lines of the subjective-probabilistic behavior displayed by the subjects of the experiments described in Sec. \ref{SecExperiments}.}
   then 
%the line of reasoning constrains 
the values that these organisms attach to decisions are shown to be constrained to be consistent with the Born rule, so this line of reasoning may play a role in showing how the innate 
``ability for performing intuitive statistics,'' developed through  biological evolution, becomes the ``roots of later acquisition of statistical principles\cite{XuGarcia08}.''
\item 
Zurek's envariance approach \cite{Zurek03,Zurek05} uses  a symmetry in the physics of entangled quantum states of two systems to argue that  a property pertaining to just one of the systems and associated with basis states of the system is  constrained to have values given by the Born rule.  This is of course suggestive of and consistent with the property so constrained indeed being objective probability given by the Born rule.  Now objective probability plays certain roles; e.g.,   the smallness of the objective probability  of an event is sufficient  to explain why this event does not occur, or occurs rarely.\footnote{Kolmogorov: ``We apply the theory of probability to the actual world of experiments in the following manner\ldots If P($A$\/) is very small, one can be practically certain that when conditions [for a repeatable experiment] are realized only once, the event $A$\/ would not occur at all \cite[pp. 3-4]{Kolmogorov56}.''}
%{\bf Cite Shafer \& Vovk?} 
But, in the envariance approach, no argument is given that the property associated with a particular basis state being small  should be taken to mean that an observer can  discount the possibility of the system existing in that state.%{\bf Or refer to discussion of objective probability comment on Papineau below}.  %So here too, without an objective basis for an explanation, probabilistic phenomena are not explained.

\item 
Vaidman has introduced an approach based on the idea of ignorance-based or self-locating uncertainty\cite{Vaidman98}, the uncertainty experienced by an experimenter after she has performed a quantum experiment that leads to Everett splitting but before she has become aware of the result.  A subjective probability is associated with this uncertainty.  McQueen and Vaidman \cite[Sec. 4.2]{McQueenVaidman19} present a thought experiment and  symmetry arguments, distinct from those employed in either the decision-theory or envariance approaches, to argue that this subjective probability must have the Born-rule value.\footnote{Kent\cite{Kent15}  and McQueen and Vaidman\cite{McQueenVaidman19}  criticize the related approach of Sebens and Carroll\cite{SebensCarroll18}  on the grounds that the latter consider  Everett copying of an observer to occur even before the observer has interacted with a device measuring a quantum system in a superposition.  We note that the viewpoint taken by Sebens and Carroll is contrary to that motivated by the Heisenberg-picture Everett ontology. See e.g. the discussion of  ``Siamese $(\nu+1)$\/-tuplets'' in \cite[Sec. 5]{Rubin03}.} Again, because only subjective probability exists in the theory, it shares with the decision-theory and envariance approaches the inability to provide explanations for physical 
phenomena.\footnote{And shares with them  as well the potential relevance for further investigations of how more sophisticated subjective  
 %judgements 
 %understanding 
 evaluations 
% the statistics 
of ordinary probabilistic phenomena develop starting from their primitive innate roots.}%subjective judgements.}
%and which leads to has occurred  
%{\bf Footnote on how McQ \&V objection to Sebens \& Carroll consistent with HP ontology?}  

\end{itemize}

On the other hand, to the extent that preclusion is taken to be an objective physical process, Everett quantum mechanics with preclusion can  provide an explanation for physical  phenomena, including but not limited to the biological evolution that may be the basis of subjective probability.%---and, in particular, what I termed above the ``empirical Principal Principle,'' the connection between subjective probability and physical law as embodied in the Born rule. 
     (The fact that preclusion involves the quantum-mechanical weight, i.e., has the Born-rule value, is  of course an {\em input}\/, whereas the other theories described above present arguments %represented 
{\em deriving}\/  %have the virtue that 
the relation between subjective probability and the Born-rule value.)% is derived.) 

The theory thus resolves the interpretational issues facing probability\cite{SEPprobability}
(with or without any connection to quantum mechanics with any number of outcomes).  It makes use of the strength of subjective interpretations of probability: regarding ordinary probabilistic phenomena, there is no question about ``what kinds of things are probabilities\cite{SEPprobability},'' they are states in the brains of organisms.  But its foundation of objective preclusion provides what purely subjective interpretations lack, the ability to explain physical phenomena including the evolution of subjective probability---in particular,  what I termed above (Sec. \ref{SecConnection}) the ``empirical Principal Principle,'' the connection between subjective probability and physical law as embodied by the Born rule. 

What Everett quantum mechanics with preclusion lacks is a {\em normative}\/   component. The consistency of the reactions of infants to the outcomes of lottery-like games with the expected statistics of those outcomes may indeed be explained by the theory. But were a precocious infant to ask ``Why {\em should}\/  I react this way? I thought probability is supposed to be `applicable,'  a `guide to life'\cite{SEPprobability}'', we could only respond by saying ``You evolved that way.''  However, as pointed out by Papineau\cite[Secs. 6-9]{Papineau10}, other interpretations of objective probability, both propensity-based and frequency-based,  are equally incapable of answering this question of ``why.''   It may be possible to add to the theory, from the outside, a normative element.  Alternatively, it may not be the case that ``the inability\dots   to provide this sort of [normative] explanation is  a deficiency.  Rather it may simply be the elimination of an illusion which, like absolute simultaneity, has no objective correlate in the physical world \cite{Rubin10}.''

%and  Sebens and Carroll rely on the concept of self-locating uncertainty\ldots  {\bf START HERE}

%%\item 
%Papineau \cite{Papineau10} argues that one can simply adjoin to the Everett interpretation the assertion that each branch of the Everett multiverse has an objective probability associated with it, with value given by the Born rule.  Having ``partnered'' with objective probability, Everett quantum mechanics can then of course explain all the phenomena that single-world quantum mechanics can. Objective probability in single-world quantum mechanics is primitive and irreducible,\footnote{We do not consider hidden-variables versions.}  and is essentially by definition an explanation for physical phenomena. Why does the ice cube placed in the glass of water not become  colder? The probability that it will become colder is extremely small. Why, after many generations, {\bf START HERE}

%\end{itemize}

%{\bf Do I really want to get into a digression on the extent to which traditional probability, can explain things, Cournot's principle, etc.? Nah\ldots But I think I'll have to because if I want to be honest I have to mention Papineau's approach}
%In single-world quantum mechanics,  reduction of the wavepacket \cite{dEspagnat76} is an an objective physical process, so the probability associated with it is objective probability.   As such it can explain the nonoccurrence of extraordinary probabilistic phenomena in the usual way that probability, explicitly or implicitly, provides explanation; i.e., via Cournot's principle, which roughly says that 

%{\bf Comment on normative issue?}

Finally, as alluded to above, there is a second motivation for  believing in the existence of preclusion, one 
completely unrelated to any role that may be claimed for preclusion  
%to be  an objeccase for the existence of preclusion can in fact be made in complete disregard of 
%any role it may play 
as a desideratum for addressing issues of probability in the Everett interpretation.   % (although it will in the end return to play that role). 
It  has been argued for decades, using a variety of approaches, that quantum mechanics and relativity taken together imply the existence of a minimum spatial length scale at or near the Planck length (for a review see \cite{Hossenfelder13}).   Buniy, Hsu and Zee \cite{BHZ05} argue  that the existence of a minimum spatial length  leads in turn to the existence of a minimum length for state vectors in quantum-mechanical Hilbert space\footnote{We note that arguments that lead to this conclusion may imply a more complicated rule for preclusion than the one employed in this paper, with a fixed $\eps_{1}$\/ in all situations.  For example, a minimum length of the Planck length $l_P$\/ suggests that a device of size $d$\/ can measure angles no smaller than $\theta_{\rm min}\approx l_P/d$\/.    The change in the state vector of a qubit caused by a rotation, which change is  itself a state vector, should be precluded if the magnitude of the rotation is smaller than  $\theta_{\rm min}$\/.  Performing a $\theta_{\rm min}$\/ rotation about the $y$\/-axis of a state $\left(\begin{array}{c}1\\0 \end{array}\right)$\/ spin-up along the $z$\/-axis, and assuming $d \gg l_P$\/ 
%---quite likely, given the smallness of  $l_P$\/---
  so $\theta_{\rm min} \ll 1$\/,  the norm squared of the difference between the original and rotated states is approximately $\left | \left(\begin{array}{c}1\\0 \end{array}\right)-\left(1-\frac{i}{2}\theta_{\rm min}\left(\begin{array}{c c}0 & -i \\ i & 0 \end{array}\right)\right)\left(\begin{array}{c}1\\0 \end{array}\right)\right |^2=(l_P/2d)^2.$\/    }---that is, to preclusion of states with norm below some minimum value. 

So, whether or not we find preclusion desirable for its ability to explain probability in Everett quantum mechanics, preclusion may well be present in quantum physics desired or not. Given its presence, it will play a role vis-\`{a}-vis probabilistic phenomena, as investigated in BHZ, \cite{Rubin10}, \cite{Hsu20} and the present paper.  Probability may thus be a telltale sign of quantum gravity making its effect known in our daily lives.

\section*{Acknowledgments}

I would like to thank Jianbin Mao, Jacob A. Rubin and Allen J. Tino for helpful discussions, and Rainer Plaga and Michael Price for comments on an earlier paper on this subject.
I am particularly grateful to  Michael Weissman for  pointing out to me 
the possibility of modifying the rule linking mathematical formalism to probability while retaining unchanged
the structure of the underlying formalism.


\begin{thebibliography}{99}

\bibitem{Everett57}H. Everett~III, `` `Relative state' formulation of quantum 
mechanics,'' {\em Rev. Mod. Phys.}\/ {\bf 29}  454-462 (1957). Reprinted in \cite{DeWittGraham73}.

\bibitem{DeWittGraham73}B.~S.~DeWitt and 
N.~Graham, eds., {\em The Many Worlds Interpretation of Quantum Mechanics}
(Princeton University Press, Princeton, NJ, 1973). 

\bibitem{Price95}  M.~C.~Price, ``The Everett FAQ,'' https://www.hedweb.com/everett/ (1995). 

\bibitem{Barrett99}J.~A.~Barrett, {\em The quantum mechanics of minds and worlds}\/
(Oxford University Press, Oxford,1999). 


\bibitem{Vaidman02} L.~Vaidman, ``Many-worlds interpretation of quantum mechanics,'' {\em The Stanford Encyclopedia of Philosophy}\/ (Fall 2018 Edition), E.~N.~Zalta, ed., https://plato.stanford.edu/archives/fall2018/entries/qm-manyworlds/.


\bibitem{HewittHorsman09}C.~Hewitt-Horsman, ``An introduction to many worlds in quantum computation,'' {\em Found. Phys.}\/ {\bf 39},
826-902 (2009); arXiv:0802.2504.

\bibitem{Saundersetal10}S.~Saunders, J.~Barrett, A.~Kent and D.~Wallace, eds., {\em Many Worlds? Everett, Quantum Theory \& Reality}\/, (Oxford University Press, Oxford, 2010).

\bibitem{Wallace12}D.~Wallace,  {\em The Emergent Multiverse: Quantum Theory According to the Everett Interpretation}\/, (Oxford University Press, Oxford, 2012).

\bibitem{Wallace13}D.~Wallace, ``The Everett interpretation,'' in R.~Batterman, ed., {\em The Oxford Handbook of Philosophy of Physics}\/ 
(Oxford University Press, Oxford, 2013).

\bibitem{DeWitt72} B.~S.~DeWitt, ``The many-universes interpretation of
quantum mechanics,'' in {\em Proceedings of the International
School of Physics ``Enrico Fermi'' Course IL: Foundations of Quantum Mechanics},\/ (Academic
Press, Inc., New York, 1972). Reprinted in \cite{DeWittGraham73}

\bibitem{Graham73}N.~Graham, ``The measurement of relative frequency,'' in \cite{DeWittGraham73}.

\bibitem{Okhuwa93}Y.~Okhuwa, ``Decoherence functional and probability interpretation,''
{\em Phys. Rev. }\/ {\bf D48}, 1781-1784 (1993).

\bibitem{Kent90} A.~Kent, ``Against many-worlds interpretations,'' {\em Int. J. Mod. Phys.}\/ {\bf A5} 1745 (1990); arXiv:gr-qc/9703089.

\bibitem{Deutsch99}D.~Deutsch, ``Quantum theory of probability and decisions,'' {\em Proc.
Roy. Soc. Lond. A}\/ {\bf 455}, 3129-3137 (1999); quant-ph/9906015.

\bibitem{Barnumetal00}H.~Barnum, C.~M.~Caves, J.~Finkelstein, C.~A.~Fuchs and R.~Schack, ``Quantum probability from decision theory?'' {\em Proc. Roy. Soc. Lond.} {\bf A456}, 1175-1182 (2000); arXiv:quant-ph/9907024


\bibitem{Rae09}A.~I.~M.~Rae, ``Everett and the Born rule,'' {\em Studies in History and Philosophy of Modern Physics} {\bf 40} 243-250 
(2009); arXiv:0810.2657.

\bibitem{Saunders10} S.~Saunders, ``Chance in the Everett interpretation,'' in \cite{Saundersetal10}, 181-205.

\bibitem{Papineau10} D.~Papineau, ``A fair deal for Everettians,'' in  \cite{Saundersetal10}, 206-226.

\bibitem{Wallace10} D.~Wallace, ``How to prove the Born rule,'' in  \cite{Saundersetal10}, 227-263.

\bibitem{GreavesMyrvold10} H.~Greaves and W.~Myrvold, ``Everett and evidence,'' in  \cite{Saundersetal10}, 264-304.

\bibitem{Kent10} A.~Kent, ``One world versus many: The inadequacy of Everettian accounts of evolution, probability, and scientific confirmation,''  in  \cite{Saundersetal10}, 307-334.

\bibitem{Albert10} D.~Albert, ``Probability in the Everett interpretation,''  in  \cite{Saundersetal10}, 355-368.

\bibitem{Price10} H.~Price, ``Decisions, decisions, decisions: Can Savage salvage Everettian probability?,''  in  \cite{Saundersetal10}, 369-390.

\bibitem{Zurek10} W.~H.~Zurek, ``Quantum jumps, Born's rule, and objective reality,'' in  \cite{Saundersetal10}, 409-432.

\bibitem{Schack10} R.~Schack, ``The principal principle and probability in the many-worlds interpretation,''  in  \cite{Saundersetal10}, 467-475.

\bibitem{Vaidman11}
L.~Vaidman
``Probability and the Many-Worlds interpretation of quantum theory, ''
in {\em Proceedings of the Conference "Quantum Theory, Reconsideration of Foundations" : V\"{a}xj\"{o} (Smaland), Sweden, 17-21 June, 2001}
(V\"{a}xj\"{o} University Press, V\"{a}xj\"{o}, Sweden, 2001)); arXiv:quant-ph/0111072

\bibitem{SebensCarroll18}C.~T.~Sebens and S.~M.~Carroll,
``Self-locating uncertainty and the origin of probability in Everettian quantum mechanics.'' {\em The British Journal for the Philosophy of Science}\/, {\bf 69}, 25-74 (2018);arXiv:1405.7577.

\bibitem{Geroch84}R.~Geroch, ``The Everett interpretation,'' {\em No\^{u}s}\/ {\bf 18}, 617-633 (1984).

\bibitem{BHZ06} R.~ V.~Buniy, S.~D.~H.~Hsu and  A.~Zee, ``Discreteness and the origin of probability in quantum mechanics,''
{\em Phys.Lett. B}\/ {\bf 640}, 219-223 (2006) ;  arXiv:hep-th/0606062.

\bibitem{Rubin10}M.~A.~Rubin, ``Explanation, evolution and subjective probability in Everett quantum mechanics with positive preclusion,'' arXiv:0912.4239. (2010). 

\bibitem{Sorkin94} R.~D.~Sorkin, ``Quantum mechanics as quantum measure theory,'' {\em Mod. Phys. Lett.}\/ {\bf A9}  3119-3128 (1994); gr-qc/9401003.

\bibitem{Sorkin97} R.~D.~Sorkin, ``Quantum measure theory and its interpretation,'' in  D.~H.~Feng and B.-L.~Hu, eds., 
{\em Quantum Classical Correspondence: Proceedings of the 4th Drexel Symposium on Quantum Nonintegrability, held Philadelphia, September 8-11, 1994}\/, 229-251 (International Press, Cambridge Mass., 1997); gr-qc/9507057.

\bibitem{Sorkin07a} R.~D.~Sorkin, ``An exercise in `anhomomorphic logic' ,''
{\em J. Phys.}\/ Conf. Ser. {\bf 67}, 012018 (2007); quant-ph/0703276.

\bibitem{Sorkin07b} R.~D.~Sorkin, ``Quantum dynamics without the wave function,'' {\em J. Phys. A} {\bf 40}, 3207-3222 (2007); quant-ph/0610204.

\bibitem{DowkerGhaziTabatabai08b} F.~Dowker and Y.~Ghazi-Tabatabai, ``Dynamical wave function collapse models in quantum measure theory,''
 {\em J. Phys. A}\/ {\bf 41}, 205306 (2008); arXiv:0712.2924.

\bibitem{DowkerGhaziTabatabai08a} F.~Dowker and Y.~Ghazi-Tabatabai, ``The Kochen-Specker theorem revisited in quantum measure theory,'' 
{\em J. Phys. A}\/ {\bf  41}, 105301 (2008); arXiv:0711.0894.

\bibitem{Barnettetal07}M.~Barnett, F.~Dowker and D.~Rideout, ``Popescu-Rohrlich boxes in quantum measure theory,''
  {\em J. Phys. A} {\bf 40},  7255-7264 (2007); quant-ph/0605253.

\bibitem{Craigetal07}D.~Craig, F.~Dowker, J.~Henson, S.~Major, D.~Rideout and R.~D.~Sorkin, ``A Bell inequality analog in quantum measure theory,''
{\em  J. Phys. A}\/ {\bf 40}, 501-523 (2007);  quant-ph/0605008.


\bibitem{Ghazi-TabatabaiWallden09a}Y.~Ghazi-Tabatabai and P.~Wallden,
``Dynamics \& predictions in the co-event interpretation,''
     {\em J. Phys. A}\/, {\bf 42}, 235303 (2009); arXiv:0901.3675.

\bibitem{Ghazi-Tabatabai09}Y.~Ghazi-Tabatabai, ``Quantum measure theory: A new interpretation,''arXiv:0906.029.
 
\bibitem{Ghazi-TabatabaiWallden09b}Y.~Ghazi-Tabatabai and P.~Wallden,
 ``The emergence of probabilities in anhomomorphic logic,''
 {\em J. Phys.}\/ Conf. Ser. {\bf 174}, 012054 (2009);  arXiv:0907.0754. 
 
\bibitem{Gudder09} S.~Gudder, ``An anholonomic logic for quantum mechanics,'' arXiv:0910.3253 (2009).
	
\bibitem{Gudder11} S.~Gudder, ``Quantum measures and the coevent interpretation,'' {\em Repts. Math.
Phys}\/  {\bf 67}, 137-156 (2011); arXiv:1005.2242.
 
 \bibitem{FraucaSorkin17} A.~M.~Frauca, R.~D.~Sorkin, ``How to measure the quantum measure,'' {\em Int. J. Theor. Phys.}\/ 
 {\bf 56}, 232?258 (2017);arXiv:1610.02087  
 


\bibitem{Galvan07a} B.~Galvan, ``Typicality vs. probability in trajectory-based formulations of quantum mechanics,''
{\em Found. Phys.} {\bf 37}, 1540-1562 (2007); arXiv:quant-ph/0605162.

\bibitem{Galvan07b} B.~Galvan, ``Origin of which-way information and generalization of the Born rule,'' 
arXiv:0705.2877.


\bibitem{Galvan08a} B.~Galvan,  ``Quantum mechanics and imprecise probability,''  {\em J. Stat. Phys.}\/ {\bf 131}, 1155-1167
(2008); arXiv:0711.3000.



\bibitem{Galvan08b} B.~Galvan, ``Generalization of the Born rule,''
{\em  Phys. Rev. A}\/ {\bf 78}, 042113 (2008); arXiv:0806.4935.




\bibitem{Hanson03} 
  R.~D.~Hanson, ``When worlds collide: Quantum probability from observer selection?,''
{\em Found. Phys.} {\bf 33}, 1129-1150 (2003);  quant-ph/0108070.  

\bibitem{Hanson06}R.~D.~Hanson, ``Drift-diffusion in mangled worlds quantum mechanics,''
{\em Proc. Roy. Soc. A}\/,, {\bf 462},  1619-1627 (2006); quant-ph/0303114.




\bibitem{Cournot1843} A.~A.~Cournot, {\em Exposition de la th\'{e}orie des chances et des probabilit\'{e}s}\/  (Hachette,
Paris, 1843). Reprinted as Volume I of B.~Bru ed., {\em Antoine-Augustin Cournot, Oeuvres
compl\`{e}tes}\/ (Vrin, Paris, 1984).


\bibitem{ShaferVovk06}G.~Shafer and V.~Vovk, ``The sources of Kolmogorov's {\em Grundbegriffe},'' {\em Statistical Science} {\bf 21},
70-98 (2006).




\bibitem{Kolmogorov56}A.~N.~Kolmogorov, {\em Foundations of the Theory of Probability}, 2nd English ed. (Chelsea Publishing Company, New York, 1956).



\bibitem{Rubin03}M.~A.~Rubin, ``Relative frequency and probability in the Everett interpretation of Heisenberg-picture quantum mechanics,'' {\em Found. Phys.} {\bf 33}  379-405 (2003); 
quant-ph/0209055.

\bibitem{dEspagnat76}  B.~d'Espagnat, {\em Conceptual Foundations of Quantum Mechanics}\/,
2nd edn. (W.~A.~Benjamin, Reading, MA, 1976).

\bibitem{Rubin01}  M.~A.~Rubin,  ``Locality in the Everett interpretation of 
Heisenberg-picture quantum mechanics,'' {\em Found. Phys. Lett.}\/ {\bf 14}, 301-322 (2001);
quant-ph/0103079.

\bibitem{Rubin04}M.~A.~Rubin, ``There is no basis ambiguity in Everett quantum mechanics,'' 
{\em Found. Phys. Lett.}\/ {\bf 17}, 323-341 (2004); arXiv:quant-ph/0310186.

\bibitem{Rubin06}M.~A.~Rubin, ``Spatial degrees of freedom in  Everett quantum mechanics,'' {\em Found. Phys.}\/ {\bf 36}, 1115-1159 (2006);arXiv:quant-ph/0511188.

\bibitem{Beckenstein01} J.~D.~Beckenstein, ``Limitations on quantum information
from black hole physics,'' {\em Acta Phys.Polon.}\/ {\bf B32}, 3555-3570 
(2001); quant-ph/0110005.


\bibitem{Deutsch02} D.~Deutsch, ``The structure of the multiverse,'' {\em Proc. Roy. Soc. London Ser. A}\/
{\bf 458} 2911-2923 (2002); arXiv:quant-ph/0104033.

\bibitem{Stein84}H.~Stein, ``The Everett interpretation of quantum mechanics: Many worlds or none?,''
{\em No\^{u}s}\/ {\bf 18}. 635-652 (1984).

\bibitem{Farhietal89}E.~Farhi, J.~Goldstone and S.~Gutmann, ``How probability arises in quantum mechanics,'' {\em Ann. Phys.  (NY)}\/ {\bf 192},
368-382 (1989).

\bibitem{Renyi70} A.~R\'{e}nyi, {\em Foundations of Probability}\/ (Holden-Day, Inc., San Francisco, 1970).


\bibitem{Oldhametal09} K.~Oldham, J.~Myland and J.~Spanier, {\em An Atlas of Functions}\/, 2nd ed.(Springer Science+Business Media, LLC, New York, 2009).



\bibitem{deFinetti89} B.~de~Finetti, ``Probabilism: A critical essay on the theory of probability and on the value of science,'' {\em Erkenntnis}\/
{\bf 31}, 169-223 (1989).

\bibitem{Lewis80} D.~Lewis, ``A  subjectivist's guide to objective chance,'' in R.~Jeffrey, ed., 
{\em Studies in Inductive Logic and Probability, Vol. II.}\/ (University of California Press, Berkeley, 1980).


\bibitem{BrownBenPorath20}H.~R.~Brown and G.~Ben~Porath, ``Everettian probabilities, the Deutsch-Wallace theorem and the Principal Principle,''
in
M.~Hemmo  and O.~Shenker, eds.,
{\em Quantum, Probability, Logic:
The Work and Influence of Itamar Pitowsky}\/
(Springer Nature Switzerland AG, Cham, Switzerland, 2020)

\bibitem{Earman18}J.~Earman, ``The Relation between credence and chance: Lewis' `Principal Principle' is a theorem of quantum probability theory,'' http://philsci-archive.pitt.edu/id/eprint/14822 (2018). 

\bibitem{Davies74}P.~C.~W.~Davies, {\em The Physics of Time Asymmetry} (University of California Press, Berkeley, 1974).

\bibitem{Teglasetal07}E.~T\'{e}glas, V.~Girotto, M.~Gonzalez and L.~Bonatti, ``Intuitions of probabilities shape expectations about the future at 12 months and beyond,'' {\em PNAS} \/{\bf 104}19156-19159 (2007).


\bibitem{TeglasBonatti16}E.~T\'{e}glas and L.~Bonatti, ``Infants anticipate probabilistic but not deterministic outcomes,'' {\em Cognition}\/, {\bf 157} 227-236 (2016).

\bibitem{XuGarcia08}F.~Xu and V.~Garcia, ``Intuitive statistics by 8-month-old infants,'' {\em PNAS}\/ {\bf 105}, 5012-5025 (2008).

\bibitem{HanusCall14}D.~Hanus and J.~Call ``When maths trumps logic: probabilistic judgements in chimpanzees,'' {\em Biology Letters}\/
{\bf 10},  20140892 (2014).

\bibitem{Rakoczyetal14} H.~Rakoczy, A.~Cl\"{u}ver, L.~Sauke, N.~Stoffregen, A.~Gr\"{a}bener, J.~Migura and J.~Call, ``Apes are
intuitive statisticians,'' {\em Cognition} {\bf 131} 60-68 (2014).

\bibitem{Tecwynetal17} E.~C.~Tecwyn, S.~Denison, E.~J.~E.~Messer and D.~Buchsbaum, ``Intuitive probabilistic inference in capuchin monkeys,'' {\em Anim. Cogn.}\/ {\bf 20} 243-256 (2017).

\bibitem{DePetrilloRosati19}F.~De~Petrillo and A.~G.~Rosati, ``Rhesus macaques use probabilities to predict future events,'' 
{\em Evolution and Human Behavior}\/ {\bf 40} 436-446 (2019).









\bibitem{DenisonXu19}S.~Denison and F.~Xu, ``Infant statisticians: The origins of reasoning under uncertainty,'' {\em Perspectives on Psychological Science}\/ {\bf 14}, 499-509 (2019).


\bibitem{deFinetti17} B.~de Finetti, {\em Theory of Probability: A Critical Introductory Treatment}\/ (John Wiley \& Sons Ltd., West Sussex, 2017)

\bibitem{Koopmans40}B.~O.~Koopmans, ``The axioms and algebra of intuitive probability,'' {\em Annals of Mathematics}\/, 2nd series, {\bf 41}, 269-292 (1940).
\bibitem{Forrester07} A.~Forrester, ``Decision theory and information propagation in quantum physics,'' {\em History and Philosophy of Science Part B: Studies In History and Philosophy of Modern Physics}\/ {\bf 38}, 815-831 (2007); arXiv:quant-ph/0604133.
%
%\bibitem{SebensCarroll18}C.~T.~Sebens and S.~M.~Carroll, ``Self-locating uncertainty and the
%origin of probability in Everettian quantum mechanics,''
%{\em Brit. J. Phil. Sci.}\/ {\bf 69},  25-74 (2018).
 


\bibitem{DeutschHayden00} D.~Deutsch and P.~Hayden, ``Information flow in entangled quantum systems,'' {\em Proc. R. Soc. Lond.}\/ {\bf A.456}, 1759-1774 (2000);arXiv:quant-ph/9906007. 

\bibitem{Rubin02}M.~A.~Rubin, ``Locality in the Everett Interpretation of Quantum Field Theory,''
{\em Found.Phys.} {bf 32}, 1495-1523 (2002); arXiv:quant-ph/0204024.


\bibitem{HewittHorsmanVedral07a}C.~Hewitt-Horsman and V.~Vedral, ``Developing the Deutsch-Hayden approach to quantum mechanics,'' {\em New Journal of Physics}\/ {\bf 9} 135 (2007); arXiv:quant-ph/0609085.

\bibitem{HewittHorsmanVedral07b} C.~Hewitt-Horsman and  V.~Vedral, ``Entanglement without nonlocality,'' {\em Phys. Rev.}\/  {\bf A 76}, 062319 (2007); arXiv:quant-ph/0611237.

\bibitem{Rubin11}M.~A.~Rubin, ``Observers and locality in Everett quantum field theory,'' 
{\em Found.Phys.}\/ {\bf 41}, 1236-1262 (2011);arXiv:0909.2673.

\bibitem{Deutsch12} D.~Deutsch, ``Vindication of quantum locality,'' {\em Proc. R. Soc. A}\/ {\bf 468}, 531-544 (2012); arXiv:1109.6223.

\bibitem{Bedard20}C.~A.~B\'{e}dard, ``Topics on quantum locality," arXiv:2001.09646 (2020).

\bibitem{KuypersDeutsch20}S.~Kuypers and D.~Deutsch, ``Everettian relative states in the Heisenberg picture,'' arXiv:2008.02328 (2020). 

\bibitem{Heisenberg58}W.~Heisenberg,  {\em Physics and Philosophy}\/ (Prometheus Books, Amherst, 1958).

\bibitem{Heisenberg74}W.~Heisenberg, {\em Across the Frontiers}\/, (Harper and Row, Publishers, Inc., New York 1974).

\bibitem{Fleming92}G.~Fleming, ``The actualizion of potentialities in contemporary quantum theory,'' {\em The Journal of Speculative Philosophy}\/ {\bf 6} 259-276 (1992).

\bibitem{Pangle14}T.~Pangle, ``On Heisenberg's key statement concerning ontology,'' {\em The Review of Metaphysics}\/ {\bf 67}, 835-859 (2014).

\bibitem{Bschir16}K.~Bschir, ``Potentiality in natural philosophy,'' in T.~E.~Eastman, M.~Epperson and D.~R.~Griffin, eds.,  
{\em Physics and Speculative Philosophy: Potentiality in Modern Science}\/ (De Gruyter, Berlin, 2016).

\bibitem{Jaeger17}G.~Jaeger, ``Quantum potentiality revisited,'' {\em Phil. Trans. R. Soc. A}\/ {\bf 375}, 20160390 (2017).

\bibitem{Jaeger17b}G.~Jaeger, `` `Wave-packet reduction' and the quantum character of the actualization of potentia,'' {\em Entropy}\/
{\bf 19}, 513 (2017).

\bibitem{BishopBrenner17}R.~R.~Bishop and J.~E.~Brenner, ``Potentiality, actuality and non-separability in quantum and classical physics: Res potentiae in the macroscopic world,'' arXiv:1801.01471 (2017).

\bibitem{Kastner18} R.~E.~Kastner, ``Taking Heisenberg's potentia seriously,'' {\em Int. J. Quant. Found.}\/ {\bf 4}, 158-172 (2018).

\bibitem{Kistler18}M.~Kistler, ``Potentiality in physics,'' in K.~Engelhard and M.~Quante, eds.,  {\em Potentiality in Physics}\/
(Springer, Dordrech, 2018).

\bibitem{Sanders18}G.~Sanders, ``An Aristotelian approach to quantum mechanics,'' https://www.academia.edu/download/56222749/An\_Aristotelian\_Approach\_to\_ Quantum\_Mechanics.pdf  (2018).


\bibitem{DeWitt98}B.~DeWitt, ``The quantum mechanics of isolated systems,'' {\em Int. J. Mod. Phys. A}\/ {\bf 13}, 1881-1916 (1998).

\bibitem{Polley00} L.~Polley, ``Quantum-mechanical probability from the symmetries of two-state systems,''   quant-ph/9906124 (2000).

\bibitem{Wallace02}D.~Wallace, ``Quantum probability and decision theory, revisited,'' arXiv:quant-ph/0211104 (2002).

\bibitem{Wallace03} D.~Wallace, ``Everettian rationality: defending Deutsch's approach
to probability in the Everett interpretation,'' {\em Stud. 
Hist. Phil. Mod. Phys.}\/ {\bf 34}, 415-439 (2003); quant-ph/0303050.

\bibitem{Wallace07}D.~Wallace, ``Quantum probability from subjective likelihood: Improving
on Deutsch's proof of the probability rule,''{\em Stud. 
Hist. Phil. Mod. Phys.}\/  {\bf 38}, 311-332 (2007); quant-ph/0312157.



\bibitem{DeWitt70}B.~S.~DeWitt, ``Quantum mechanics and reality,'' {\em Physics Today}\/ {\bf 23}, 30-35 (1970).

\bibitem{VanEsch07}P.~Van ~Esch, ``On the Born rule and the Everett programme,''
{\em Annales de la Fondation Louis de Broglie}\/ {\bf 32}, 51-59  (2007),

\bibitem{Barrett17}J.~A.~Barrett, ``Typical worlds,'' {\em Studies in History and Philosophy of Science Part B: Studies in History and Philosophy of Modern Physics}\/
{\bf 58}, 31-40 (2017); arXiv:1912.05312

\bibitem{Vaidman20}L.~Vaidman,
``Derivations of the Born rule,'' in
M.~Hemmo  and O.~Shenker, eds.,
{\em Quantum, Probability, Logic:
The Work and Influence of Itamar Pitowsky}\/
(Springer Nature Switzerland AG, Cham, Switzerland, 2020)

\bibitem{Page82}D.~N.~Page, ``The Einstein-Podolsky-Rosen physical reality is completely described by quantum mechanics,''  {\em Phys. Lett.} {\bf  A91}, 57-60 (1982).

\bibitem{Stapp85} H.~P.~Stapp, ``Bell's theorem and the foundations of quantum physics,''
{\em Am. J. Phys.}\/ {\bf 53}, 306 (1985).

\bibitem{Tipler00} F.~J.~Tipler, ``Does quantum nonlocality exist? Bell's theorem and the many-worlds interpretation,'' arXiv:quant-ph/0003146 (2000).

\bibitem{Bacciagaluppi02} G.~Bacciagaluppi, ``Remarks on space-time and locality in Everett's interpretation.,'' in  
T.~Placek and J.~Butterfield, {\em Non-locality and Modality}\/ (Kluwer Academic Publishers, Dordrecht, 2002); philsci-archive.pitt.edu/id/eprint/504. 

\bibitem{TimpsonBrown02} C.~G.~Timpson and  H.~R.~Brown, ``Entanglement and relativity,'' in   R.~Lupacchini and V.~Fano,  eds., {\em Understanding Physical Knowledge}\/ (CLUEB, Bologna, 2002).

\bibitem{Vaidman15} L.~Vaidman, ``The Bell inequality and the many-worlds interpretation,'' in M.~Bell and
S.~Gao, eds., {\em Quantum Nonlocality and Reality: 
50 Years of Bell's Theorem}\/ (Cambridge University Press, Cambridge, 2106);
arXiv:1501.02691


\bibitem{Bell04} J.~S.~Bell, {\em Speakable and Unspeakable in Quantum Mechanics}\/, 2nd ed. (Cambridge University Press, Cambridge, 2004).

\bibitem{Bell64}J.~S~Bell, ``On the Einstein-Podolsky-Rosen paradox,'' reprinted in \cite{Bell04} (1964).

\bibitem{Bell75}J.~S~Bell, ``The theory of local beables,'' reprinted in \cite{Bell04} (1975). 

\bibitem{Bell81}J.~S~Bell,  ``Bertlmann's socks and the nature of reality,'' reprinted in \cite{Bell04} (1981).

\bibitem{Bell90}J.~S~Bell, ``La nouvelle cuisine,'' reprinted in \cite{Bell04} (1990).


\bibitem{Norsen06}T.~Norsen, ``Bell locality and the nonlocal character of nature,''
{\em Found. Phys. Lett.} {\bf 19}, 633-655 (2006); arXiv:quant-ph/0601205.


\bibitem{Blaylock10}G.~Blaylock, ``The EPR paradox, Bell's inequality, and the question of locality,''
{\em Am. J. Phys.}\/ {\bf 78}, 111-120 (2010);arXiv:0902.3827.

\bibitem{Maudlin14} T.~Maudlin, ``What Bell did,''  {\em J. Phys. A: Math. Theor.}\/ {\bf 47} 424010 (2014); arXiv:1408.1826.

\bibitem{Myrvoldetal20} W.~Myrvold, M.~Genovese and A.~Shimony, ``Bell's Theorem,'' {\em The Stanford Encyclopedia of Philosophy }\/(Fall 2020 Edition), Edward N. Zalta ed., https://plato.stanford.edu/archives/fall2020/entries/bell-theorem/.

\bibitem{Weissman99}M.~B.~Weissman, ``Emergent measure-dependent probabilities from modified quantum dynamics without state-vector reduction,''
{\em Found. Phys. Lett.}\/ {\bf 12}  407-426   (1999); arXiv:quant-ph/9906127.

\bibitem{Deutsch16} D.~Deutsch, ``The logic of experimental tests, particularly of Everettian quantum theory,''  {\em Studies in History and Philosophy of Modern Physics}\/ {\bf 55} 24-33 (2016).

\bibitem{Zurek03} W.~H.~Zurek, ``Environment-assisted invariance, entanglement, and probabilities in quantum physics,'' 
{\em Phys. Rev. Lett.}\/ {\bf 90} 120404 (2003); arXiv:quant-ph/0211037.

\bibitem{Zurek05} W.~H.~Zurek, ``Probabilities from entanglement, Born's rule 
$p_k=|\psi_k|^2$\/ from envariance,'' {\em Phys. Rev. A}\/ {\bf 71} 052105  (2005); arXiv:quant-ph/0405161.

\bibitem{Vaidman98} L.~Vaidman,  ``On schizophrenic experiences of the neutron or why we should believe in the many-worlds interpretation of quantum theory,''  {\em International Studies in the Philosophy of Science}\/ {\bf 12} 245-261 (1998); arXiv:quant-ph/9609006.


\bibitem{McQueenVaidman19} K.~J.~McQueen and L.~Vaidman, ``In defence of the self-location uncertainty account of probability in the many-worlds interpretation,'' {\em Studies in History and Philosophy of Science Part B: Studies in History and Philosophy of Modern Physics}\/
{\bf 66} 14-23 (2019); 

\bibitem{Kent15}A.~Kent, ``Does it make sense to speak of self-locating uncertainty in the universal wave function? Remarks on Sebens and Carroll,'' {\em Foundations of Physics} {\bf 45} 211?217 (2015); arXiv:1408.1944.
%
%\bibitem{SebensCarroll18}C.~T.~Sebens and S.~M.~Carroll, ``Self-locating uncertainty and the
%origin of probability in Everettian
%quantum mechanics,''
%{\em Brit. J. Phil. Sci.}\/ {\bf 69} 25?74 (2018);arXiv:1405.7577 

\bibitem{SEPprobability} A.~H\'{a}jek, ``Interpretations of Probability,'' {\em The Stanford Encyclopedia of Philosophy}\/ (Fall 2019 Edition), 
E.~ N.~Zalta ed., https://plato.stanford.edu/archives/fall2019/entries/probability-interpret/.

\bibitem{Hossenfelder13} S.~Hossenfelder,
``Minimal length scale scenarios for quantum gravity,''
{\em Living Rev. Relativity},\/ {\bf 16}, 2 (2013).

\bibitem{BHZ05} R.~ V.~Buniy, S.~D.~H.~Hsu and  A.~Zee, ``Is Hilbert space discrete?'' 
{\em Phys. Lett. B}\/ {\bf 630}, 68-72 (2005); arXiv:hep-th/0508039.

\bibitem{Hsu20} S.~D.~H.~Hsu,  ``Discrete Hilbert space, the Born rule, and quantum gravity,'' arXiv:2007.12938 (2020).






\end{thebibliography}
\end{document}